\documentclass{emulateapj}
\usepackage{natbib}
\usepackage{epsfig}
\usepackage{graphicx}
\usepackage{subfigure}
\usepackage{float}
\usepackage{amsmath}
\usepackage{color}
\usepackage{amssymb}
\usepackage{amsfonts}

\usepackage[colorlinks,linkcolor=blue,anchorcolor=green,citecolor=blue]{hyperref}
\usepackage{amssymb}
\usepackage{upgreek}
\usepackage[normalem]{ulem}
\usepackage{verbatim}

\def\be{\begin{equation}}
\def\ee{\end{equation}}

\usepackage{CJK}
\usepackage{upgreek}
\usepackage{booktabs}

\shorttitle{Magnetospheric curvature radiation as FRBs}
\shortauthors{Wang et al.}
\begin{document}

\title{Magnetospheric curvature radiation by bunches as emission mechanism for repeating fast radio bursts}


\author{Wei-Yang Wang \altaffilmark{1,2,3}, Yuan-Pei Yang \altaffilmark{4}, Chen-Hui Niu \altaffilmark{3}, Renxin Xu \altaffilmark{1,2}, Bing Zhang \altaffilmark{5}}
\affil{$^1$School of Physics and State Key Laboratory of Nuclear Physics and Technology, Peking University, Beijing 100871, P.R.China}\email{wywang@bao.ac.cn}
\affil{$^2$Kavli Institute for Astronomy and Astrophysics, Peking University, Beijing 100871, P.R.China}\email{r.x.xu@pku.edu.cn}
\affil{$^3$National Astronomical Observatories, Chinese Academy of Sciences, Beijing 100101, P.R.China}
\affil{$^4$South-Western Institute for Astronomy Research, Yunnan University, Kunming, Yunnan, 650500, P.R.China}\email{ypyang@ynu.edu.cn}
\affil{$^4$Department of Physics and Astronomy, University of Nevada, Las Vegas, NV 89154, USA}

\begin{abstract}
Coherent curvature radiation as the radiation mechanism for fast radio bursts (FRBs) has been discussed since FRBs were discovered.
We study the spectral and polarization properties of repeating FRBs within the framework of coherent curvature radiation by charged bunches in the magnetosphere of a highly magnetized neutron star.
The spectra can be generally characterized by multisegment broken power laws, and evolve as bunches move and the line of sight sweeps.
Emitted waves are highly linear polarized if the line of sight is confined to the beam within an angle of $1/\gamma$, while circular polarized degree becomes strong for off-beam cases.
The spectro-temporal pulse-to-pulse properties can be a natural consequence due to the magnetospheric geometry.
We investigate the relationship between drift rate, central frequency and their temporal duration.
The radius-to-frequency mapping is derived and simulated within the assumptions of both dipolar and quadrupolar magnetic configurations.
The geometric results show that FRBs are emitted in field lines more curved than open field lines for a dipolar geometry.
This suggests that there are most likely existing multipolar magnetic configurations in the emission region.
\end{abstract}

\keywords{Radio bursts (1339); Radio transient sources (2008); Radiative processes (2055); Neutron stars (1108); Magnetars (992)}

\section{Introduction}\label{sec1}

The field of fast radio bursts (FRBs,  \citealt{Lorimer07}) enjoyed a rapid development from both the observational and theoretical frontiers (see \citealt{Cordes19,Petroff19,Zhang20}, for reviews).
Up to the present, there are hundreds of FRB sources been discovered\footnote{See Transient Name Server, \href{https://www.wis-tns.org/}{https://www.wis-tns.org/}.}, and dozens of them can repeat (e.g., \citealt{CHIME21}).
Although the large number of FRBs have been detected, the physical origin(s) of FRBs are still unknown.
The discovery of a million Jansky FRB-like burst from a Galactic magnetar SGR J1935+2154 \citep{Bochenek20,CHIME20b} suggests that magnetars are the most likely candidate sources for at least some FRBs.

Regarding the nature of FRBs, observationally, the brightness temperature of FRBs can reach $\sim10^{35}$ K, which demands an extremely coherent radiation mechanism.
\cite{Melrose17} summarized three forms of coherence: radiation by bunches, reactive instabilities (or plasma masers), and kinetic instabilities (or vacuum masers).
Most FRB models invoke neutron stars or magnetars to explain the milliseconds duration, high luminosity, and GHz emission frequency (e.g., see \citealt{Platts19}, for a review).
Based on the distance of the emission region from the neutron star, these models can be generally divided into two categories \citep{Zhang20}: pulsar-like models (emission within the magnetosphere of a compact object, e.g., \citet{Kumar20,Wang20,Lu20,Yang21}), and gamma-ray-burst-like models (emission from a relativistic shocks far outside the magnetospheres, e.g., \citet{Metzger19,Beloborodov20,Margalit20}).
There exist some challenges from both the observational \citep{Luo20,Li21,Nimmo21} and theoretical \citep{Lu18,Lu20} aspects for the latter models.
In this paper, we mainly consider one type of pulsar-like models, which focuses on coherent curvature radiation by charged bunches, to demonstrate its application to repeating FRBs\footnote{It is an open question that whether all FRBs repeat. Observationally FRBs fall into two groups: repeaters and apparently non-repeating FRBs. We mainly focus on repeaters in this paper.}.

Polarization measurements are potential tools to reveal more original information of FRBs. Most FRBs usually have strong linear polarization, which is generally dozens of percent, with some even close to 100\% (e.g., \citealt{Michilli18,Day20,Luo20}), but some also have significant circular polarization fractions \citep{Hilmarsson21b,Kumar21,Xu21}.
These properties are similar to those of pulsars, which show a wide variety of polarization fractions between sources, but with noticeable differences.
There is at least a fraction of pulsars whose S-shaped polarization angle swing can be characterized by the rotation vector model \citep{Lorimer12}.
However, some FRBs (FRBs 121102, 180916) exhibit flat polarization angle (PA) across each pulse \citep{Michilli18,Nimmo21}, some others (e.g. FRBs 180301, 181112) show variable PAs across each burst \citep{Cho20,Luo20}, and FRB 20201124A shows both flat and evolving PA \citep{Hilmarsson21b,Kumar21}.

FRBs have been reported in detection from 110 MHz to at least 8 GHz \citep{Pastor20,Gajjar18}.
The spectral indices are volatile, ranging from $-10$ to $+14$ if they are modeled by simple power laws \citep{Spitler16}.
This may imply a narrow-band spectrum with intrinsic hardening spectral index at low frequencies. 
Within the framework of coherent curvature radiation by bunches, 
\cite{Yang20} proposed that charge separation between the electron and positron clumps can explain the narrow-band spectrum.

Notably, a very intriguing spectral pattern has been found in at least some repeaters.
These repeaters exhibit a clear time--frequency downward drifting, that is, the central frequency of consecutive sub-bursts with later-arriving time have lower frequencies \citep{CHIME19a,CHIME19b,Hessels19,Josephy19,CHIME20a,Fonseca20,Day20,Li21,Platts21}.
The downward drifting structure is nicknamed the ``sad trombone'' effect. However, rarely there also exists possible upward-drifting event (the ``happy trombone'' effect) for some bursts \citep{Chawla20,CHIME20a,Hilmarsson21a,Kumar21}. The two peaks of the FRB-like event from SGR J1935+2154 seem to be consistent with such a pattern \citep{Bochenek20,CHIME20b}.

There have been several models proposed to explain these spectro-temporal characteristics. Within the framework of magnetospheric curvature radiation, we proposed a generic geometric model to explain the downward drifting pattern \citep{Wang19}. Invoking possible time difference in launching the sparks, we show that a small fraction of FRBs can show an upward drifting pattern \citep{Wang20}. 
Other mechanisms intrinsic to the FRB sources include magnetar-wind-driven external shock \citep{Metzger19}, asteroid falling \citep{Liu20}, blast waves from flares \citep{Beloborodov20}, relativistic moving source \citep{Rajabi20}, and free-free absorption \citep{Kundu21}, which have been proposed to interpret the downward drifting features. Models invoking propagation effects, e.g. plasma lensing \citep{Cordes17} or scintillation \citep{Simard20}, can account for both upward and downward drifting patterns but are challenged by the fact that the majority of dirfting patterns are downdrift. 

In this paper, we attempt to interpret the spectral and polarization properties of FRBs within the framework of coherent curvature radiation by bunches in the magnetosphere of a magnetar. The pulse-to-pulse properties are explained by the geometric model and we confront the model with the repeater data.
The paper is organized as follows.
We discuss the dynamics of moving charges and their radiation properties in Section \ref{sec2}.
The pulse-to-pulse properties are demonstrated and applied to derive radius-to-frequency mapping in Section \ref{sec3}.
The results are discussed and summarized in Section \ref{sec4}.
The convention $Q_x=Q/10^x$ in cgs units is used throughout the paper.

\section{Burst from moving charges in the magnetosphere}\label{sec2}

In contrast to the continuous ``sparking'' processes in the polar cap region of normal pulsars that produce particle bunches \citep{RS75}, FRB emissions are believed to be triggered by a sudden and violent mechanism.
Here, we do not discuss the trigger mechanism, but attempt to study the radiation properties from the bunched particles that likely formed during the violent triggering process.
For a relativistic charged particle in a strong magnetic field, the vertical momentum perpendicular to the field line drops to zero in rapidly, so that its trajectory essentially tracks with the magnetic field lines.
When these charged particles stream outward along the magnetic field lines and enter the charge starvation region, they are accelerated by the electric field parallel to the B-field. A photon-pair production cascade may be triggered, which mimic the sparking processes in inner magnetosphere of radio pulsars but with a much larger amplitude. 

Basically, electron-positron pairs are created by such a sparking mechanism, move along the curved trajectories, and emit curvature radiation due to the perpendicular acceleration.
Curvature radiation from charged bunches has been invoked to interpret both coherent radio emission of pulsars (e.g., \citealt{RS75,Sturrock75,Elsaesser76,Cheng77,Melikidze00,Gil04,Gangadhara21}) and FRBs (e.g., \citealt{Katz14,Kumar17,Yang18,Ghisellini18,Katz18,Lu18,Wang20,WangLai20,Cooper21}).

As the bunches flow out along the field lines, they co-rotate with the neutron star.
The observed duration of an FRB due to the possible rotation is 
\be
w\sim\frac{\min (\theta_{\rm jet} P / 2 \pi,\,t_{\rm int})}{1+z},
\label{eq:duration}
\ee
where $\theta_{\rm jet}$ is the burst ``jet'' beaming angle, $P$ is the period, $z$ is the redshift and $t_{\rm int}$ is the intrinsic duration of the FRB \citep{Yang19}.

\subsection{Charged particles in the magnetosphere}\label{sec2.1}
 
\begin{figure}
\includegraphics[width=0.48\textwidth]{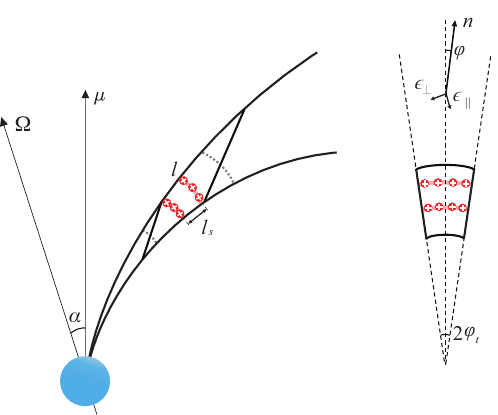}
\caption{\small{ Schematic diagram of a bulk of bunches contributing to instantaneous
radiation. In the left panel, within the moving bulk, the light solid red line shows the slice in which charges emit with roughly the same phase. Therefore, two red lines present two bunches. The bunch length is $l$ with the mean space between each bunch $l_s$ in the laboratory frame. The grey dotted lines denote average heights for both upper and lower boundaries of the bulk. The bulk seen in horizon plane is shown in the right panel. The unit vector of line of sight (LOS) is denoted by $\boldsymbol{n}$, and $\boldsymbol{\epsilon}_{\|}$ and $\boldsymbol{\epsilon}_{\perp}$ denote the two polarization components (see Appendix \ref{app:coherent}). The opening angle of the bulk of the emission region is $2\varphi_t$.}}
\label{fig:charges}
\end{figure}

A neutron star may undergo a sudden trigger (e.g. crust cracking), which sustains at least milliseconds. Such a process likely creates many sudden and violent sparking events.
These sparking events may generate bunches via two-stream instabilities \citep{Usov87,Asseo98,Melikidze00}.
Bunches are numerically simulated to formed in open field lines through a cascading process \citep{Philippov20}.
\cite{Benacek21} found the overlap of fastest particles with consecutively emitted bunches in the momentum space via Particle-in-Cell  simulations. 
This overlap drives two-stream instabilities, causing up to $\sim15\%$ of the initial kinetic energy transformed into electric field energy and eventually to radiation. 

When the size of the charged bunch is smaller than the half wavelength, waves are coherently enhanced significantly.
However, the charges moving in different trajectories (different distances to the triggering source) produce different photon arrival delays, which leads to incoherence of radiation.
Charges in the bunch which are projected in the horizontal plane would have the same phase.
As shown in Figure \ref{fig:charges}, for instance, the charges\footnote{Positive net charges are the point of interest in the following discussion.} within a light red slice (a bunch) are created at the same time and travel the same distance, emitting electromagnetic waves that are added coherently.
The emission from a charge is coherently added within one bunch but incoherently bunch to bunch if $l_s>\lambda/2$.

Emitting bunched particles are most likely produced in the open field lines rather than closed field lines, where there may be absorption caused by bunches moving along adjacent field lines \citep[e.g.][]{Yang18}.
The magnetic field is much stronger near the polar cap region, leading to possibly more dramatic activities.
Multipolar magnetic fields may exist near the stellar surface and may be dominant at at least several tens of stellar radii.
It makes field lines more curved and equivalently enlarge the polar cap angle of the open field lines.
In following calculations, the spherical coordinates ($r,\,\theta,\,\varphi$) with respect to the magnetic axis are used.

Consider generally geometric expressions of the magnetic field configuration.
We discuss the simplest case where the field lines are axisymmetric (see Appendix \ref{app:geometry}).
Combine with the Maxwell equations and the force-free condition, a magnetic field configuration solution is given by Equations (\ref{eq:B-field}) and (\ref{eq:Bgeo}).
The parameter $n$ denotes the order of the multipoles, e.g., $n=1$ for dipole, $n=2$ for quadrupole, etc.
Regardless of the value of $n$, some self-similar features for multipolar configuration can be summarized as follows:
\begin{enumerate}
    \item The trajectory family with the same polar angle $\theta$ has the same tangent direction;
    \item Curvature radius is only proportional to $r$ for the trajectory family with the same polar angle $\theta$, see Equation (\ref{eq:curvatureradius});
    \item $\theta_{\rm max}$ is smaller for a high-$n$ multipolar configuration due to more curved field lines;
    \item The magnetic field strength is generally in the form of $B\simeq B_s (R/r)^{2n+1}$, where $B_s$ is the field strength at the stellar surface and $R$ is the stellar radius;
    \item For any $\theta\ll1$, the distance that charges traveled along the field lines is $s\approx r$.
\end{enumerate}

A generic expression for the volume of bunches in the lab-frame is given by (see Appendix \ref{app:geometry})
\be
V_b\simeq \lambda\varphi_t r^2 \frac{f'(\theta_s)}{f'(\theta)}\Delta \theta_s,
\label{eq:volume}
\ee
where $\theta_s$ is the angle of the footpoint for each field line at the stellar surface.
The term of $f'(\theta_s)/f'(\theta)$ depends on $n$.
For small $\theta$ values, the magnetic field lines are not curved enough, so that $f'(\theta_s)/f'(\theta)$ is roughly of the order of unity.

For an anti-parallel rotator, the polar cap region is populated with net positive changes. Positrons produce radiation as the outflow streams along the curved B-field lines.
However, the typical cooling timescale is much shorter than the FRB duration, so that there must be an electric field $E_{\|}$ parallel to the $B$-field to sustain energy of charged particles \citep{Kumar17}.
The $E_{\|}$ enable pairs to decouple and the radiation would not cancel out \citep{Yang20}.
Generally, the acceleration of a charged bunch can be described as
\be
N_e E_{\|} e d s-\mathcal{L}_b d t=N_e m_{e} c^{2} d \gamma,
\label{eq:accelaration}
\ee
where $N_e$ is the number of net charge in one bunch that are radiating coherently, $c$ is the speed of light, $e$ is elementary charge, $m_{e}$ is the positron mass, $\gamma$ is the Lorentz factor of bunch and $\mathcal{L}_b$ is the luminosity of the radiation.

Charges in the bunch are suggested to move along nearly identical orbits, therefore they act like a single macro-charge, which emits a power $N_e^2$ times $p_e$, where $p_e$ is the curvature radiation luminosity for one positron, which reads $2e^2\gamma^4c/(3\rho^2)$, where $\rho$ is the curvature radius.
As shown in Figure \ref{fig:charges}, there could be $N_b$ bunches contributing to instantaneous radiation in the bulk, but the emissions from them are added incoherently, leading to an incoherent summation of the curvature radiation power, i.e., $\mathcal{L}=N_bN_e^2p_e$, where $N_b$ is the number of independent bunches that contribute to the observed luminosity at an epoch.
The number $N_b$ can be estimated as $\sim 10^5 r_7\gamma_2^{-1}\theta_{-1}^{-1}$ (emission within a conal angle of $1/\gamma$ can be seen at $\omega=\omega_c$, see Section \ref{sec2.3}).

The the observed isotropic luminosity should be compared withe the isotropic equivalent luminosity from the model, which is given by \citep{Kumar17}
\be
\mathcal{L}_{\rm iso}\approx\gamma^4N_bN_e^2p_e\approx4.6\times10^{42}\gamma^8_2N_{b,5}N_{e,22}^2\rho_7^{-2}\,\rm{erg\,s^{-1}},
\label{eq:Liso}
\ee
Consequently, the radiation power is always balanced with the electric power provided by the $E_{\|}$
\be
E_{\|} e v_e\simeq N_ep_e,
\label{eq:Ebalance}
\ee
where $v_e$ is the velocity of positron.
Regardless of the initial energy distribution of the charges, they will eventually be modulated by the $E_{\|}$ in the magnetosphere.
According to Equation (\ref{eq:Ebalance}), the required $E_{\|}$ to sustain the radiation power is given by
\be
E_{\|}\simeq\frac{2\gamma^4 N_e e}{3\rho^2}\simeq3.2\times10^{6}\gamma^4_2N_{e,22}\rho_7^{-2}\,\rm esu.
\label{eq:E}
\ee

The timescale for the radiation power be balanced by $E_{\|}$ is essentially the cooling timescale of the bunches, which can be estimated as
\be
t_c \approx\frac{N_e\gamma m_ec^2}{\mathcal{L}_b}\approx1.8\times10^{-12}\rho_7^2\gamma_2^{-3}N_{e,22}^{-1}\,\rm s.
\label{eq:tcooling}
\ee
This timescale is much shorter than the FRB duration.
This suggests that particles essentially stay balanced throughout the FRB emission process. 

At lower heights, the volume of a bunch can be calculated as
\be
V_b\simeq6.0\times10^{10}\varphi_{t,-2}\nu_9^{-1}r_7^2 \Delta\theta_{s,-3}\,\rm cm^3.
\ee
In the lab-frame, the number density of net charge can be estimated as
\be
n_e=\frac{N_e}{V_b}\simeq1.7\times10^{11}N_{e,22}\varphi_{t,-2}^{-1}\nu_9r_7^{-2}\Delta\theta_{s,-3}^{-1}\,\rm cm^{-3}.
\label{eq:ne}
\ee
Therefore, the required multiplicity is
\be
\begin{aligned}
\mathcal{M}&=\frac{n_e}{n_{\rm GJ}}\\
&\simeq2.4\times10^{2n-2}N_{e,22}P_0\varphi_{t,-2}^{-1}\nu_9B_{s,15}^{-1}r_7^{2n-1}\Delta\theta_{s,-3}^{-1},
\end{aligned}
\label{eq:Multiplicity}
\ee
where the Goldreich-Julian density is \citep{GJ69}
\be
n_{\rm GJ}=\frac{|\boldsymbol{\Omega}\cdot\boldsymbol{B}|}{2\pi ce},
\label{eq:nGJ}
\ee
and $\Omega$ is angle frequency of a neutron star.
The slow rotation can let $\mathcal{M}>1$ \citep{Cooper21b}.

\subsection{Breakout of FRB waves}\label{sec2.2}

An important requirement for FRBs to be observed is the successful escape of radio waves from the magnetosphere.
Near the emission region, even the electric field of the electromagnetic wave $E$ is very large, the electron motion is confined to be along magnetic field lines.
The oscillation of charge particles would not be relativistic in the presence of the electromagnetic wave.
Compared to the vacuum case, the speed of them is suppressed by a factor of $(\omega/\omega_B)^2$, where $\omega_B$ is the cyclotron frequency:
\be
\omega_B=\frac{eB}{m_ec}\simeq1.8\times10^{21-2n}B_{s,15}r_7^{-2n-1}\,\rm{rad\,s^{-1}}.
\label{eq:cyclotronfrequency}
\ee
A strength parameter can be defined as $a=eE/m_ec\omega$.
The electric amplitude can be calculated as $E\approx8.2\times10^8\mathcal{L}_{42}^{1/2}r_7$ esu, leading to $a=2.3\times10^6\mathcal{L}_{42}^{1/2}r_7\nu^{-1}_9$.
As the wave propagates away from the magnetosphere in a decreasing background field $B\propto r^{-2n-1}$, $E$ would exceed the background field at $r_0$:
\be
r_0=\left(B_sR^{2n+1}\sqrt{\frac{c}{2\mathcal{L}}}\right)^{1/2n}\approx(1.2\times10^5B_{s,15})^{1/2n}R.
\label{eq:r0}
\ee
For instance, $r_0\approx3.5\times10^2R$ for a dipole and $r_0\approx18.7R$ for a quadrupole.

When the FRB propagates in the magnetosphere, the non-linear effect due to relativistic oscillation of electrons under the coherent electromagnetic wave is involved \citep{Yang20b}.
For X mode photons\footnote{The two polarization states of electromagnetic radiation in the presence of a strong magnetic field can be described as those in which the electric vector is perpendicular to the magnetic and wave vector plane (X mode) and those with electric vector in that plane (O mode).} with $E$ much smaller than the background magnetic field $B$, the electron cross section is
\be
\sigma \simeq \left\{
\begin{aligned}
&(\omega/\omega_B)^2\sigma_T,~& \omega \ll \omega_{B} \\
&\sigma_T,~& \omega \gg \omega_{B}
\end{aligned}.
\right.
\ee
Beyond the region $r\gtrsim10^8~{\rm cm}$ in a dipolar field, the wave amplitude is stronger than the background magnetic strength, leading to the possibility that the magnetosphere is optically thick to FRBs, i.e., $\tau_{sc}\simeq n_era^2\sigma\gg1$, due to the enhanced scattering \citep{Beloborodov21}.

However, even if the magnetosphere is opaque to the FRB, the FRB can break out from the magnetosphere medium once its radiation pressure exceeds the plasma pressure in the magnetosphere, $P_{\rm rad}\gg P_{\rm gas}$.
The radiation pressure caused by the bunches is 
\be
P_{\rm rad}=\frac{\mathcal{L}}{4\pi r^2c}\simeq2.7\times10^{16}\mathcal{L}_{42}r_7^{-2}\,\rm dyn\,cm^{-2}.
\label{eq:Prad}
\ee
The pressure of the pair gas can be estimated as $P_{\rm gas}\simeq n_{\rm gas}k_B T_{\rm gas}$, where $n_{\rm gas}$ is number density of gas, $k_B$ is the Boltzmann constant and $T_{\rm gas}$ is the temperature of gas.
Consider that surrounding gas should not be denser and hotter than bunches, i.e., $n_{\rm gas}<n_e$ and $k_B T_{\rm gas}<\gamma m_ec^2$, one can obtain 
\be
\begin{aligned}
P_{\rm gas}&<n_e\gamma m_ec^2\\
&\simeq1.4\times10^{7}N_{e,22}\gamma_2\varphi_{t,-2}^{-1}\nu_9r_7^{-2}\Delta\theta_{s,-3}^{-1}\,\rm dyn\,cm^{-2}.
\label{eq:Pgas}
\end{aligned}
\ee
Therefore, the radiation pressure of the FRB can readily push aside and break out of the surrounding pair medium, so that both O- and X-mode radio waves may be observable (e.g., \citealt{Ioka20}).

\subsection{Spectrum}\label{sec2.3}

\begin{figure*}
\begin{center}
\includegraphics[width=0.96\textwidth]{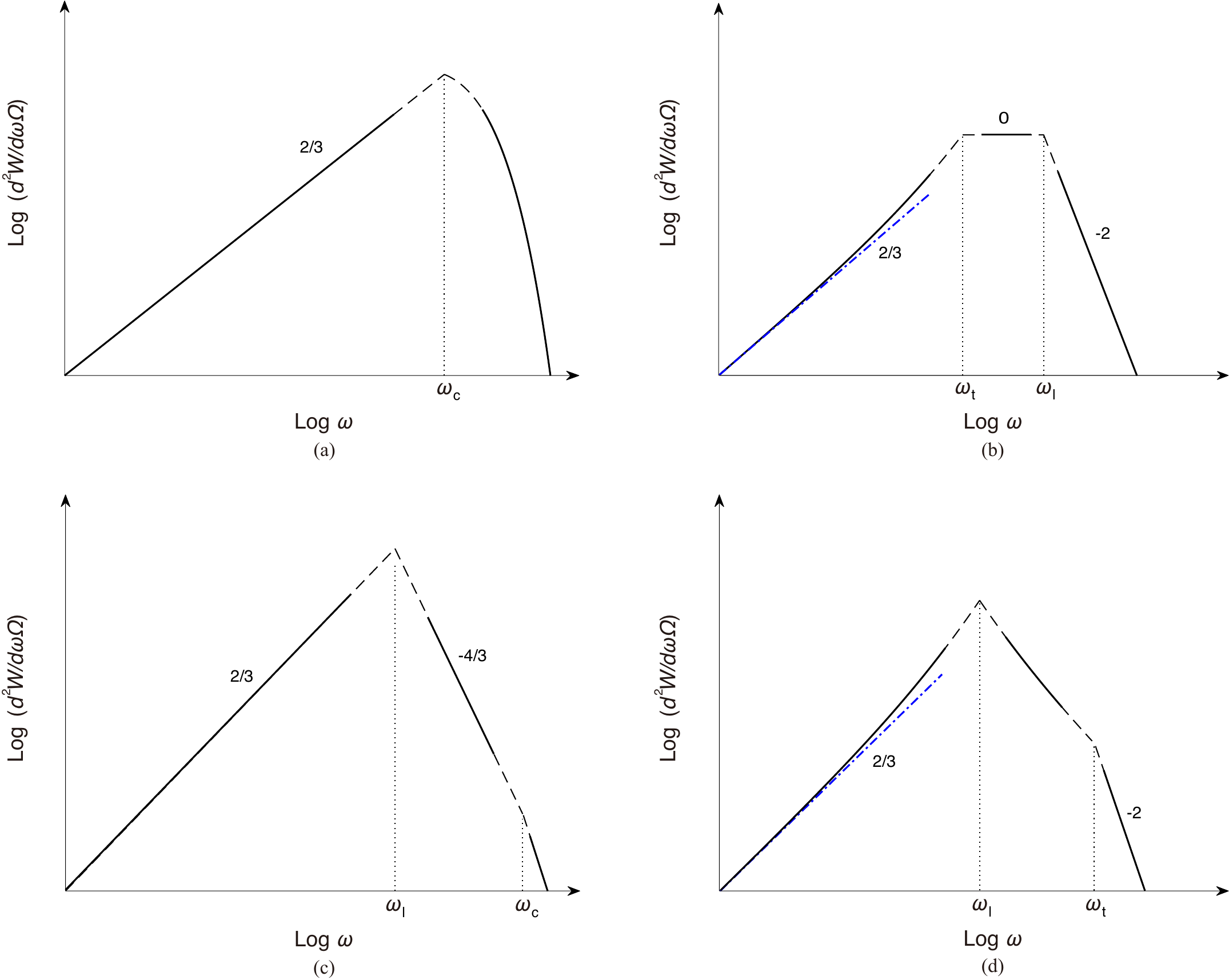}
\caption{\small{The spectra of the coherent curvature radiation: (a) $\omega_c\ll\omega_t$ and $\omega_c\ll\omega_l$; (b) $\omega_t\ll\omega_l\ll\omega_c$; (c) $\omega_l\ll\omega_c\ll\omega_t$; (d) $\omega_l\ll\omega_t\ll\omega_c$. The blue dashed-dotted lines denote power-law spectra with index of $2/3$.
}}
\label{fig:spectrum}
\end{center}
\end{figure*}


Consider the radiation from the bulk of bunches in the magnetosphere.
The observed emission cannot be simply demonstrated by the summation of the curvature radiation power of individual particles, since FRB emissions must be coherent.
The bunch length in the radial direction is $\sim10\nu_9^{-1}$ cm, much smaller
than the curvature radius of the trajectory.  This allows a point-source approximation.
The $E_{\|}$-induced acceleration is balanced by the radiation of charged particles, sustaining a constant velocity $v_e\simeq c$, and constant-$\gamma$ positron distribution.

The energy per unit time per unit solid angle is
\begin{equation}
\frac{d^2W}{d t d \Omega}=\frac{c}{4 \pi} |\boldsymbol{E}(t)\mathcal{R}|^2,
\end{equation}
where $\mathcal{R}$ is the distance from the emitting source to the observer.
The spectral information can be obtained from
Fourier transformation, leading to 
\begin{equation}
\frac{d^2W}{d\Omega d \omega}=c|\boldsymbol{E}(\omega)\mathcal{R}|^2.
\label{eq:d2I/domegadOmega}
\end{equation}
Contained in $\boldsymbol{E}(\omega)$ is all the information about the frequency behavior of $\boldsymbol{E}(t)$.
Note that this is the total energy per solid angle per frequency range in the entire pulse, and there is no ``per unit time'' in its dimension, due to the violation of the
uncertainty relation $\Delta \omega\Delta t>1$ \citep{Rybicki79}.

If the pulses repeat on an average timescale $T$, one can obtain
\begin{equation}
\frac{d^2W}{d \omega d \Omega d t} \equiv \frac{1}{T} \frac{d^2W}{d \omega d \Omega}.
\end{equation}
However, the definition of radiation power for curvature radiation from a single charge is no
longer meaningful, because the charge motion direction only sweeps the LOS once.
For coherent radiation from more than one bunch sweeping across the LOS, $T$ here would be the mean time interval of each coherent pulse, and $d^2W/d\Omega d\omega$ corresponds to the radiation energy of one coherent pulse \citep{Yang18}.

According to Equation (\ref{eq:E}), the Lorenz factor of charges strongly depends on the location in the magnetosphere due to the rapid balance between $E_{\|}$ and radiation.
A delta function distribution of the positron Lorenz factor in a macro-charged bunch is adopted, which is different from the simple power-law distribution hypothesised by \cite{Yang18}.
In this section, we assume that bunches are uniformly distributed in all directions of ($\boldsymbol{e}_r, \boldsymbol{e}_\theta, \boldsymbol{e}_\phi$) in three-dimensional bunches.
Any particle could be described by three subscripts ($i$, $j$, $k$), which contains all the information of the location.
If there is more than one charged particle, one can replace the single amplitude by the sum of the amplitudes.
The total energy radiated per unit solid angle per unit frequency interval for the moving charges in the magnetosphere can be written as
\begin{equation}
\begin{aligned}
&\frac{d^2W}{d \omega d \Omega}=\frac{e^{2} \omega^{2}}{4 \pi^{2} c}\\
&\times\left|\int_{-\infty}^{+\infty} \sum^{N_s}_{i}\sum^{N_\theta}_{j} \sum^{N_{\phi}}_{k}-\boldsymbol{\beta}_{e\perp, ijk}{\rm e}^{i \omega[t-\boldsymbol{n} \cdot \boldsymbol{r_{ijk}}(t) / c]} d t\right|^{2},
\end{aligned}
\label{eq:B1}
\end{equation}
where $N_s,\,N_{\theta},\,N_{\phi}$ are the maximum numbers for $i$, $j$, $k$, and the total number of positron is $N=N_s N_{\theta}N_{\phi}$ (see Appendix \ref{app:coherent}).

Charged particles move along different trajectories with different phases and orientations.
The frequency-dependent spread angle is found as \citep{Jackson98}
\be
\theta_{c}(\omega) \simeq \left\{
\begin{aligned}
&\frac{1}{\gamma}\left(\frac{2 \omega_{c}}{\omega}\right)^{1 / 3}=\left(\frac{3 c}{\omega \rho}\right)^{1 / 3},~& \omega \ll \omega_{c} \\
&\frac{1}{\gamma}\left(\frac{2 \omega_{c}}{3 \omega}\right)^{1 / 2},~& \omega \gg \omega_{c}
\end{aligned},
\right.
\label{eq:thetac}
\ee
where $\omega_{c}$ is the critical frequency of curvature radiation, which reads $\omega_{c}=3c\gamma^3/(2\rho)$.
For $\omega\sim\omega_c$, the conal beam angle is $\sim1/\gamma$.
The observed length of the emitting region can be estimated as $\sim r\gamma^{-1}\theta^{-1}/3\sim3.3\times10^5\rho_7\gamma_2^{-1}\theta_{-1}^{-1}$ cm, which is much smaller than $cw\sim3\times10^7$ cm.
As a result, there should be persistent bunches travel through the emitting region during the FRB emission.

In order to obtain the spectrum of  electromagnetic waves, the polarized amplitudes of the waves could be described by deriving two orthogonal components, i.e., $A_{\|}$ and $A_{\perp}$ (see Appendix \ref{app:coherent}), which contain the information not only of amplitude, but also of phase.
For a single charge, $A_{\|}$ is earlier than $A_{\perp}$ by $\pi/2$ in phase.

We calculate the amplitude of electromagnetic waves for any particle with a given trajectory (see Appendix \ref{app:coherent} for details).
The coherent condition of the trajectories with different curvature radii is satisfied, due to $\Delta \rho\sim\rho/\gamma\ll2^{2/3}\rho/3 $ for $\omega\approx\omega_c$ \citep{Yang18}.
Therefore, the total energy radiated per unit solid angle per unit frequency interval for the moving bunches is given by
\be
\frac{d^2W}{d \omega d \Omega}\simeq N_{lb}\sum_i^{N_l}\sum_j^{N_\theta}\sum_k^{N_\phi}\frac{e^{2} \omega^{2}}{4 \pi^{2} c}\left|-\boldsymbol{\epsilon}_{\|} A_{\|,ijk}+\boldsymbol{\epsilon}_{\perp} A_{\perp,ijk}\right|^2,
\ee
where $N_l$ is the number of positrons inside each bunch and $N_{lb}$ is the number of bunches.

We define a characteristic frequency
\be
\omega_{l}=\frac{2 c}{l}.
\label{eq:omega_l}
\ee
Since the mean space between charged particles increase as the bunches move away from the magnetosphere, the bunch size continuously increase with increasing height.
If $\omega\ll\omega_l$, one would have the wavelength $l<\lambda$, so that charges in the bunch emit with roughly the same phase.
Otherwise, the term $\sin^2(\omega/\omega_l)/(\omega/\omega_l)^2$ will play a role in reducing coherence (see Appendix \ref{app:coherent}).

Charges in different trajectories may have different azimuth angles.
We define
\be
\omega_\varphi=\frac{3c}{\rho(\chi'^2+\varphi'^2)^{3/2}}.
\label{eq:omega_phi}
\ee
Here, $\chi'$ and $\varphi'$ can let $\omega_\varphi$ reach the minimum value of $\omega_\varphi$, which is defined as $\omega_t$.
Note that we only investigate the cases for either $1/\gamma<\varphi'$ or $1/\gamma<\chi'$, since the total flux for both $1/\gamma<\varphi'$ and $1/\gamma<\chi'$ is extremely small.
For instance, if $\omega\ll\omega_t$\footnote{Assuming $\chi'\ll\varphi'$ here. For $\varphi'\ll\chi'$, one can replace $\varphi'$ by $\chi'$.}, the LOS is confined to the beam within $1/\gamma$.
One may define $\varphi<1/\gamma$ as on-beam and $\varphi>1/\gamma$ for off-beam.
For the on-beam case, the summation of $A_{\perp}$ is extremely small due to axial symmetry matches the calculations from \cite{Yang18}.
We develop the \cite{Yang18} model by invoking the off-beam cases, and calculate the spectra and polarization properties.

There are several regimes for the spectrum:

(a) $\omega_c\ll\omega_t$, $\omega_c\ll\omega_l$:

The spectrum is shown in Figure \ref{fig:spectrum} panel (a), and its formula can be referred to Equation (\ref{eq:spectruma}).
The spectrum could be regarded as that of a single charge multiplied by a factor of $N_e^2N_b$.
The peak energy radiated per unit solid angle per unit frequency interval is given by
\be
\left.\frac{d^2W}{d \omega d \Omega}\right|_{\rm max}\simeq N_bN_e^2\left(\frac{\chi_{u}-\chi_{d}}{\Delta\theta_s}\right)^2\frac{3e^2\gamma^2}{2^{2/3}\pi^2 c}\Gamma(2/3)^{2}.
\ee

(b) $\omega_t\ll\omega_l\ll\omega_c$:

For $\omega_t>\omega_c$, all of the radiation energy in the bunch opening angle can be observed.
The spectrum is shown in panel (b) of Figure \ref{fig:spectrum}.
For $\omega\ll\omega_l$, the spectral formula can be referred to Equation (\ref{eq:spectrumb}).
If $\omega\gg\omega_l$, the formula should be multiplied by a factor of $(\omega_l/\omega)^2$.
If $\chi\gg\varphi'$, the peak energy radiated per unit solid angle per unit frequency interval is given by
\be
\begin{aligned}
&\left.\frac{d^2W}{d \omega d \Omega}\right|_{\rm max}\simeq N_bN_e^2\left(\frac{\chi_{u}-\chi_{d}}{\Delta\theta_s}\right)^2\frac{3e^2\gamma^2}{4\pi^2c}\left(\frac{2\omega_t}{\omega_{c}}\right)^{2 / 3}\\
&\times\left[\Gamma(2 / 3)^{2}+\frac{\gamma^{2}}{4^{2/3}}\left(\frac{\chi_{u}+\chi_{d}}{2}\right)^{2} \Gamma(1 / 3)^{2}\left(\frac{\omega_t}{\omega_{c}}\right)^{2 / 3}\right].
\end{aligned}
\ee
On the other hand, if $\chi\ll\varphi'$, it  is
\be
\begin{aligned}
&\left.\frac{d^2W}{d \omega d \Omega}\right|_{\rm max}\simeq N_bN_e^2\left(\frac{\chi_{u}-\chi_{d}}{\Delta\theta_s}\right)^2\frac{3e^2\gamma^2}{4\pi^2c}\left(\frac{2\omega_t}{\omega_{c}}\right)^{2 / 3}\\
&\times\left[\Gamma(2 / 3)^{2}+\frac{\gamma^{2}}{4^{2/3}}\varphi^{2} \Gamma(1 / 3)^{2}\left(\frac{\omega_t}{\omega_{c}}\right)^{2 / 3}\right].
\end{aligned}
\ee

(c) $\omega_l\ll\omega_c\ll\omega_t$:

The case is the same as case (a) but there should be multiplied by a factor of $(\omega_l/\omega)^2$ for $\omega_l\ll\omega$.
The spectrum is shown in panel (c) of Figure \ref{fig:spectrum}.

(d) $\omega_l\ll\omega_t\ll\omega_c$:

The bunch-length-induced incoherence appears at $\omega<\omega_t$.
The spectrum is shown in panel (d) of Figure \ref{fig:spectrum}.
The peak energy radiated per unit solid angle per unit frequency interval is similar to  case (b), but one should replace $\omega_t$ by $\omega_l$.

The spectrum can evolve as the bunches flow out from the magnetosphere.
The mean space $l_s\propto r$, making high frequency waves become incoherent at higher altitudes, e.g., spectra become harder, as shown in Figure \ref{fig:spectrum}, evolve from (a) to (c), or (b) to (d).
The LOS is initially off-beam, then approaches the axis within $1/\gamma$ to become on-beam, and finally becomes off-beam again, as the LOS sweeps across the emission region.
Consequently, the spectrum evolves from (b) to (a) and then back to (b), or (d) to (c) and then back to (d).

\subsection{Polarization}\label{sec2.4}

\begin{figure*}
\begin{center}
\includegraphics[width=0.96\textwidth]{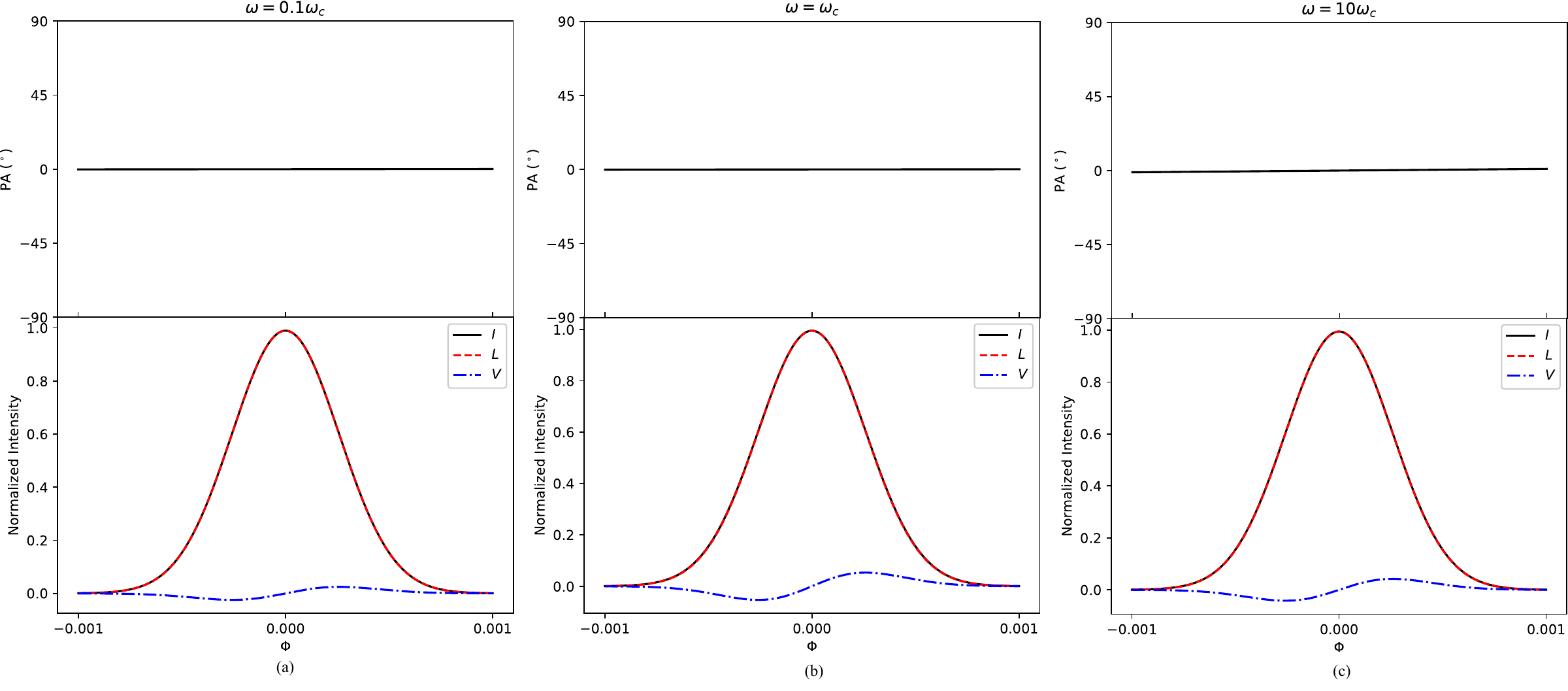}
\caption{\small{Simulated polarization profile and PA evolution for different frequencies: (a) $\omega=0.1\omega_c$; (b) $\omega=\omega_c$; (c) $\omega=10\omega_c$. Polarized intensities are plotted in colored solid lines. Parameters are adopted as $\alpha=\pi/6$, $\zeta=\pi/4$, $\gamma=100$, $\varphi_t=0.001$, $\Phi_p=0$ and $\sigma_w=\varphi_t/2$. 
}}
\label{fig:stokes1}
\end{center}
\end{figure*}

\begin{figure*}
\begin{center}
\includegraphics[width=0.96\textwidth]{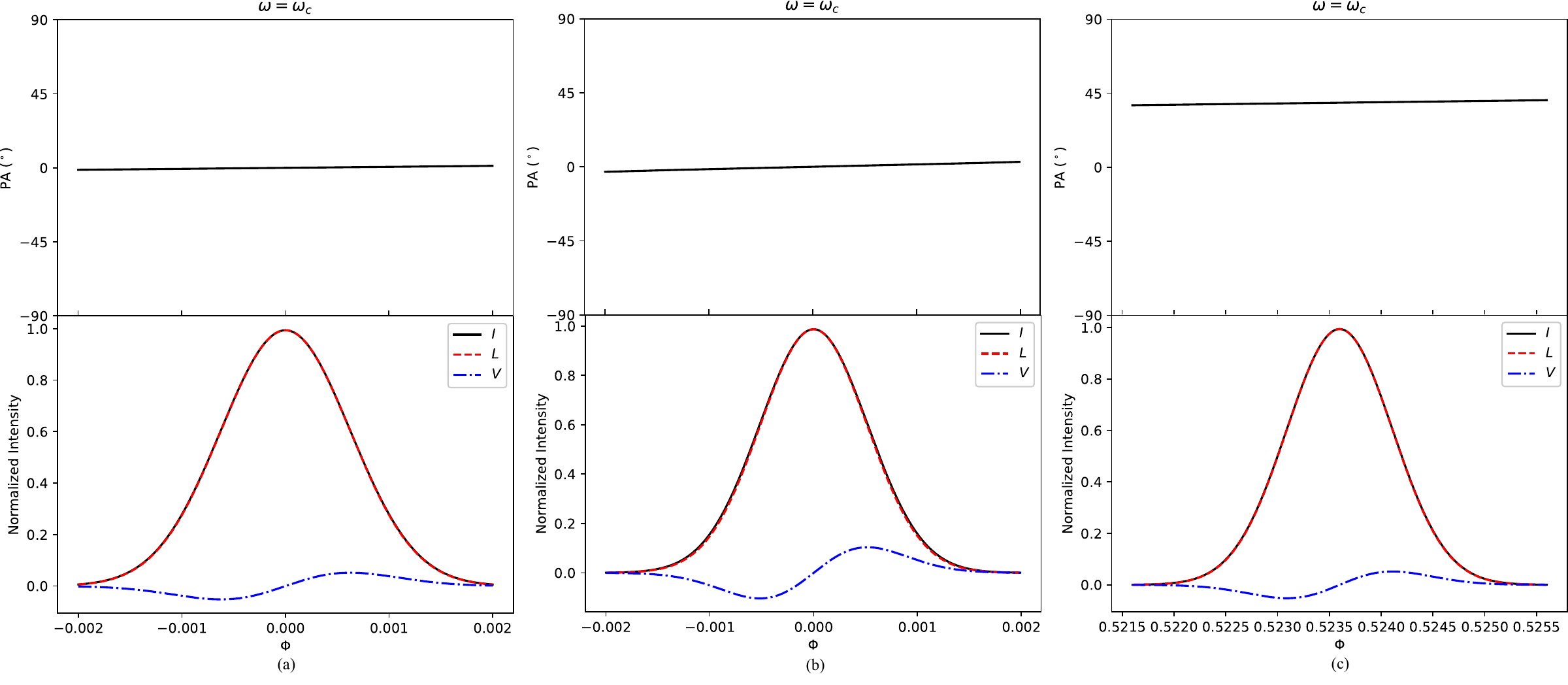}
\caption{\small{The same as panel (b) in Figure \ref{fig:stokes1} but for different parameters: (a) $\alpha=\pi/18$, $\zeta=\pi/3$, $\sigma_w=\varphi_t/2$, $\Phi_p=0$; (b) $\alpha=\pi/6$, $\zeta=\pi/4$, $\sigma_w=\varphi_t$, $\Phi_p=0$; (c) $\alpha=\pi/6$, $\zeta=\pi/4$, $\sigma_w=\varphi_t/2$, $\Phi_p=\pi/6$.
}}
\label{fig:stokes2}
\end{center}
\end{figure*}

\begin{figure*}
\begin{center}
\includegraphics[width=0.96\textwidth]{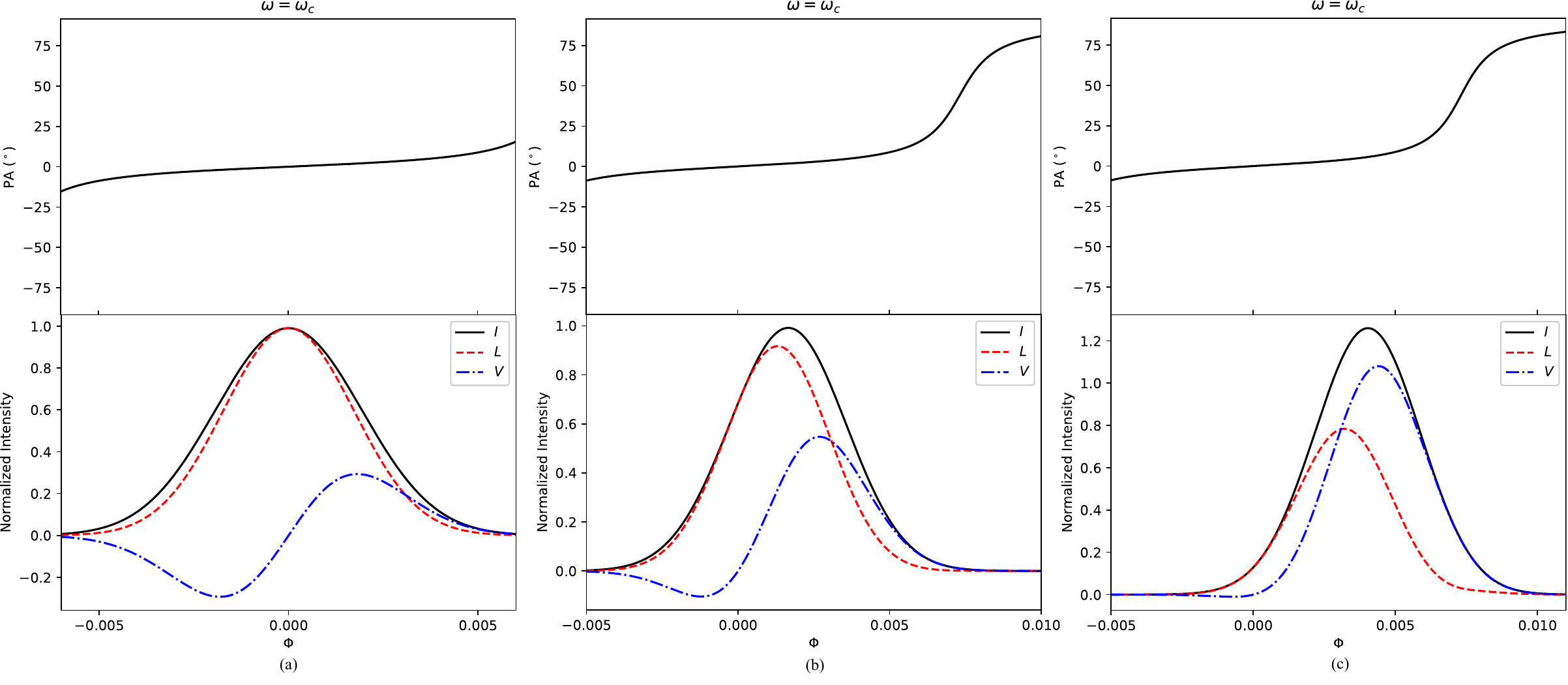}
\caption{\small{The same as panel (b) in Figure \ref{fig:stokes1} but for $\varphi_t=0.7\gamma^{-1}$ with different parameters: (a) $\Phi_{\rm p}=0$; (b) $\Phi_{\rm p}=0.002$; (c) $\Phi_{\rm p}=0.005$.
}}
\label{fig:stokes4}
\end{center}
\end{figure*}

\begin{figure*}
\begin{center}
\includegraphics[width=0.96\textwidth]{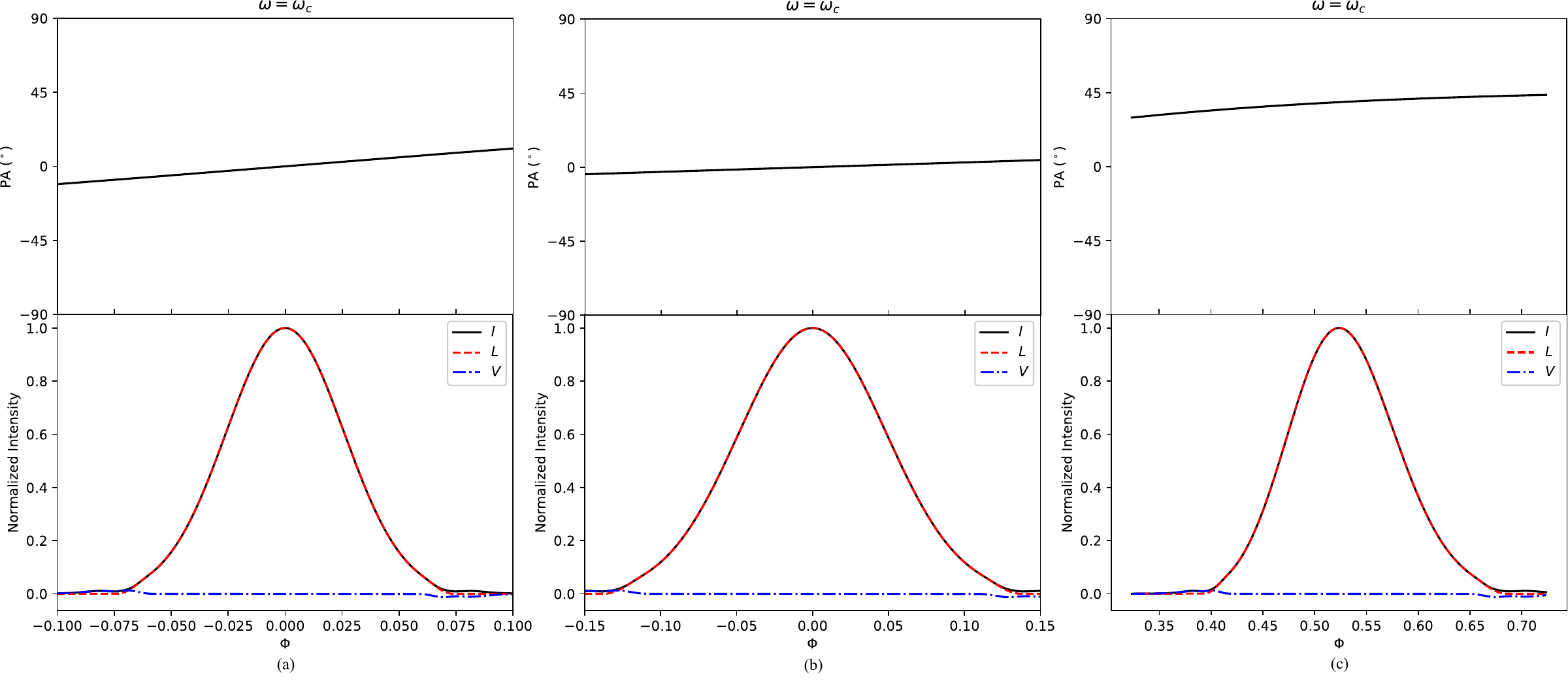}
\caption{\small{The same as panel (b) in Figure \ref{fig:stokes1} but for $\varphi_t=10\gamma^{-1}$ with different parameters: (a) $\alpha=\pi/6$, $\zeta=\pi/4$ and $\Phi_{\rm p}=0$; (b) $\alpha=\pi/18$, $\zeta=\pi/6$ and $\Phi_{\rm p}=0$; (c) $\alpha=\pi/6$, $\zeta=\pi/4$ and $\Phi_{\rm p}=\pi/6$.
}}
\label{fig:stokes3}
\end{center}
\end{figure*}

We consider the polarization of emission from the moving charged bunches.
Basically, for a single particle, if $\varphi_k=0$, the LOS would sweep in the trajectory plane, and only $A_{\|}$ can be seen.
Consequently, the radiation is 100\% linearly polarized.
Circular polarization would be seen and become stronger for off-beam observations.
However, the deviation of the LOS from the direction of the velocity makes the observed flux decrease rapidly.
It seems difficult to observe highly circularly polarized emission.

For more than one emitting particles, the polarization could be derived from the summation of the electric field vector of the waves. 
The Fourier transformed components of $\boldsymbol{E}(\omega)$ in the directions $\boldsymbol{\epsilon}_{\|}$ and $\boldsymbol{\epsilon}_{\perp}$ are given by
\be
E_{\|}(\omega)=\frac{e\omega A_{\|}}{2\pi c\mathcal{R}},\,
E_{\perp}(\omega)=\frac{e\omega A_{\perp}}{2\pi c\mathcal{R}}.
\ee
If the LOS sweeps the axis of symmetry of the emission region, i.e., $\varphi=0$, the summation of $A_{\perp}$ appears to cancel out each other, leading to 100\% linear polarization.
For off-beam observations, $\sum \boldsymbol{A}_{\perp}\not=0$, so that the linearly polarized component decreases while the circularly polarized component becomes stronger.

The Stokes parameters for observed waves can be calculated as
\begin{equation}
\begin{aligned}
&I=\mu\left(A_{\|} A_{\|}^{*}+A_{\perp} A_{\perp}^{*}\right),\\
&Q=\mu\left(A_{\|} A_{\|}^{*}-A_{\perp} A_{\perp}^{*}\right), \\
&U=\mu\left(A_{\|} A_{\perp}^{*}+A_{\perp}A_{\|}^{*}\right),\\
&V=-i\mu\left(A_{\|}A_{\perp}^{*}-A_{\perp}A_{\|}^{*}\right),
\end{aligned}
\label{eq:stokes}
\end{equation}
where $\mu=\omega^2 e^2/(4\pi^2 \mathcal{R}^2 c T)$ is the proportionality factor.
Here, the chosen way is such that $I$ is the flux density repeated on the average timescale $T$.
The corresponding PA is then given by
\be
\Psi=\frac{1}{2} \tan ^{-1}\left(\frac{U_s}{Q_s}\right),
\label{eq:PA}
\ee
where
\begin{equation}
\left(\begin{array}{l}
U_s \\
Q_s
\end{array}\right)=\left(\begin{array}{cc}
\cos 2\psi & \sin 2\psi \\
-\sin 2\psi & \cos 2\psi
\end{array}\right)\left(\begin{array}{l}
U \\
Q
\end{array}\right),
\end{equation}
in which $\psi$ is given by the rotation vector model \citep{Radhakrishnan69} as a function of azimuthal angle with respect to the spin axis $\Phi$:
\begin{equation}
\tan \psi=\frac{\sin \alpha \sin \Phi}{\cos \alpha \sin \zeta-\cos \zeta \sin \alpha \cos \Phi}.
\label{eq:RVM}
\end{equation}
When the LOS sweeps across the emission region as the neutron star spins, and if particles happen to be moving into the bulk region, pulsed emission can be seen.
Here we assume that the plasma density in the emission region is Gaussian-modulated in the azimuthal direction.
The electric field vector should be multiplied by a modulation function 
\be
f(\Phi)=f_{0} \exp \left[-\left(\frac{\Phi-\Phi_{\mathrm{p}}}{\sigma_{w}}\right)^{2}\right],
\label{eq:gaussianprofile}
\ee
where $\Phi_{\rm p}$ is the peak location of the Gaussian function, i.e., at the center of emission region, $f_0$ is the amplitude and $\sigma_w$ is the Gaussian width.

Consider a quadrupole magnetic configuration.
We assume that $\Delta\theta_s=0.002$ and $r=10^7$ cm.
Adopting $f_0 = 1$, $\sigma_w=\varphi_t/2$ and $\Phi_{\rm p} = 0$, we simulate the polarization profile and PA evolution in three frequency cases ($\omega=0.1\omega_c$, $\omega=\omega_c$, $\omega=10\omega_c$), under the assumption of $\varphi_t=10^{-3}$, $\gamma=100$, $\alpha=\pi/6$ and $\zeta=\pi/4$ as shown in Figure \ref{fig:stokes1}.
We then consider the polarization properties related to different geometric conditions.
The polarization properties are simulated with different declination angles, Gaussian widths and peak locations in panel (b) of Figure \ref{fig:stokes2}.

The radiation is highly polarized, i.e. $I^2 = L^2 + V^2$, as in the case of 100\% elliptically
polarized radiation, where $L^2=Q^2+U^2$ denotes linearly polarized component.
According to the calculation in Section \ref{sec2.3}, if $\varphi'\ll1/\gamma$ and $\chi'\ll1/\gamma$, the circular polarization degree can be estimated as $|V|/I\sim2|A_{\perp}/A_{\|}|$.
It would reach the maximum vale at $\omega\sim\omega_c$.
In Figure \ref{fig:stokes1}, $|L|/I > 94\%$
and $|V|/I < 33\%$ within the pulse window ($-\varphi_t<\varphi<\varphi_t$), suggesting a high degree of linear polarization.
The ratio $|L|/I > 98\%$ and $|V|/I < 18\%$ within the pulse window ($-\varphi_t<\varphi<\varphi_t$) in panel (a) and (c) of Figure \ref{fig:stokes2}, and $|L|/I > 94\%$ and $|V|/I < 30\%$ for panel (b) of Figure \ref{fig:stokes2}.

All figures show a flat PA evolution within the burst phases.
The PA evolves within $2^\circ$ as shown in Figure \ref{fig:stokes1}.
For different geometric conditions, PA also evolves at most within $6^\circ$ (see Figure \ref{fig:stokes2}).

The circular polarized degree becomes stronger as $\varphi_t$ gets larger.
We consider the cases that $\varphi_t=0.007$.
Particles distributed in bunches can affect the summation of amplitudes, hence, are considered with different forms.
The polarization profile and PA evolution are simulated and illustrated in Figure \ref{fig:stokes4}.
Within the pulse window, all three panels in Figure \ref{fig:stokes4} exhibit $|V/I>50\%|$.
Circular polarization can dramatically grow when the peak location of the particle distribution deviates from the geometric center. PAs evolve as S-shapes in panel (b) and (c), which is generally consistent with some observations of highly circular polarized bursts \citep{Kumar21,Xu21}.

If $\varphi\ll1/\gamma$, $A_{\|}$ would be comparable with $A_{\perp}$ when $\omega\sim\omega_t$, leading to a high degree of circular polarization.
We simulate the polarization profile and PA evolution at $\omega=\omega_c$ for $\varphi_t=0.1$ in Figure \ref{fig:stokes3}.
The parameters are adopted as follows: $\alpha=\pi/6$, $\zeta=\pi/4$ and $\Phi_{\rm p}=0$ for panel (a); $\alpha=\pi/18$, $\zeta=\pi/6$ and $\Phi_{\rm p}=0$ for panel (b); $\alpha=\pi/6$, $\zeta=\pi/4$ and $\Phi_{\rm p}=\pi/6$ for panel (c).
For all the cases, waves are highly linearly polarized but a $\sim100\%$ circular polarization exists at near $|\varphi|\simeq\varphi_t$.
However, the flux for the off-beam case is extremely lower than that for on-beam case.
The PA evolves within $20^\circ$ for panel (a) and $14^\circ$ for panel (b) and (c).
The evolution more dramatically changes with the magnetospheric geometry compared with the case of $\varphi'\ll1/\gamma$.

\section{Pulse-to-pulse properties}\label{sec3}

\subsection{Drift pattern of sub-pulse}\label{sec3.1}

\begin{figure}
\includegraphics[width=0.48\textwidth]{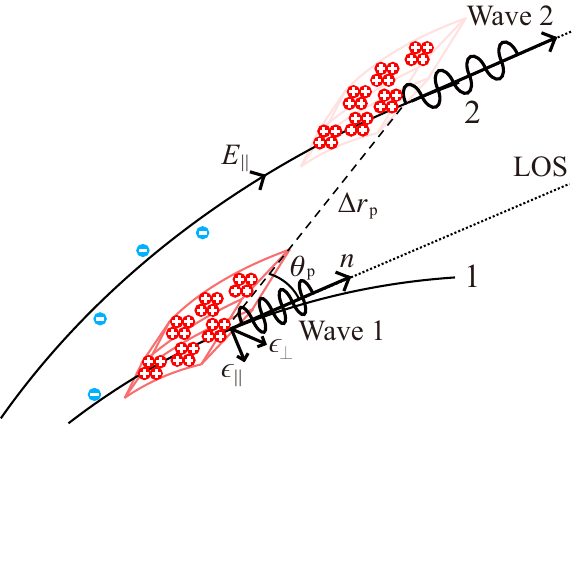}
\caption{\small{A schematic diagram of moving charges in the magnetosphere. For instance, the observer could see emission from the two bulks of neighboring magnetic field lines 1 and 2. The dotted lines show the LOS. The two emitting points for wave 1 and wave 2 are different in radius and in azimuthal angle.
}}
\label{fig:sub-pulse}
\end{figure}

\begin{figure}
\includegraphics[width=0.48\textwidth]{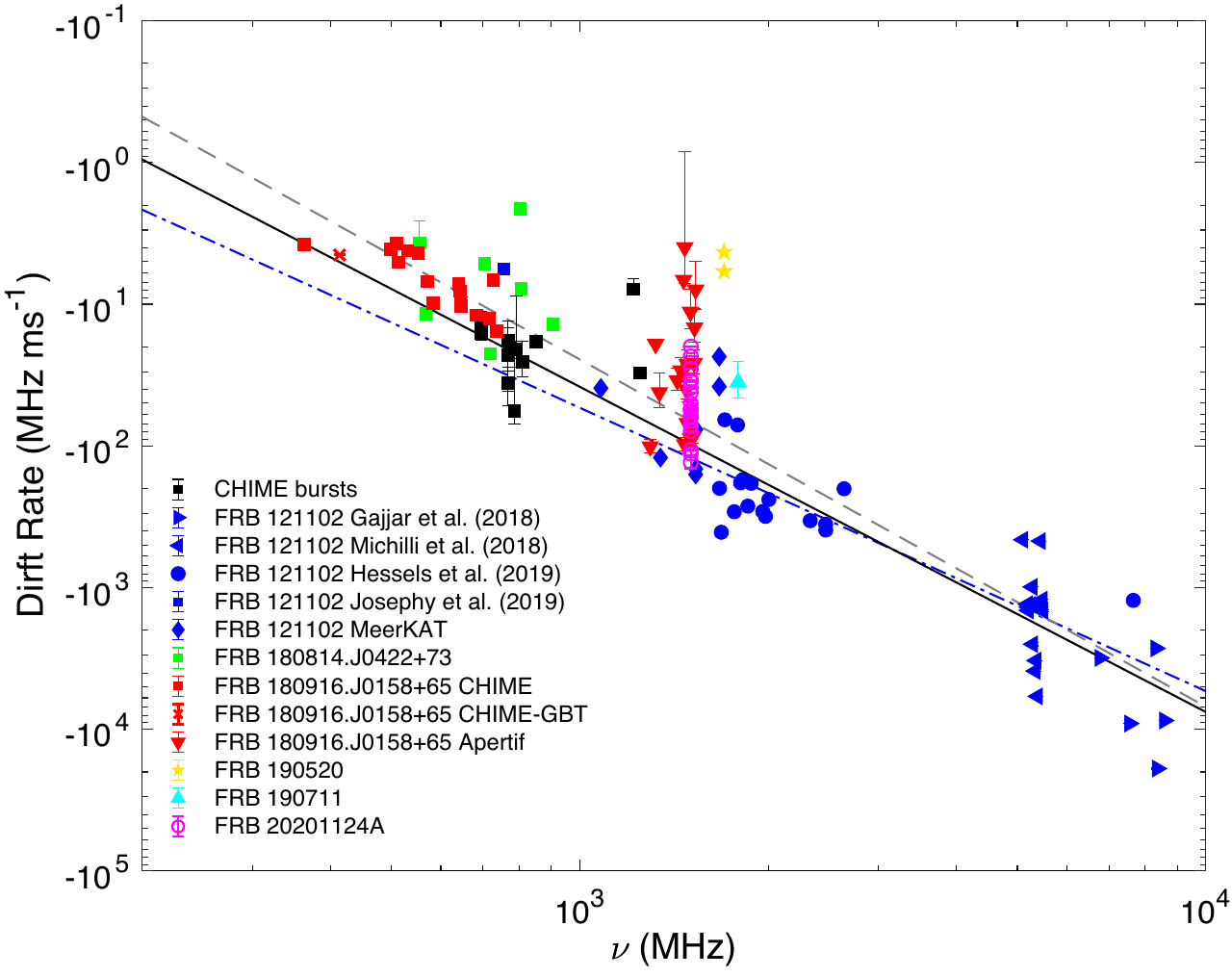}
\caption{\small{Drift rates at different frequencies.
Both drift rates and frequencies are transferred in the rest frame of the host galaxy. Different repeater are presented in different colors: FRB 121102 (blue), FRB 180814.J0422+73 (green), FRB180916.J0158+65 (red), FRB 190711 (cyan), FRB 20201124A (purple), FRB 190520 (yellow) and other CHIME bursts (black). 
The grey dashed line shows the best
fit through the data for all FRBs.
The black solid dashed line is the best fit for FRB 121102 with a power law index of 2.29 and the blue dashed-dotted line for that with a frozen index of 2.00.}}
\label{fig:ratevsnu}
\end{figure}

\begin{figure}
\includegraphics[width=0.48\textwidth]{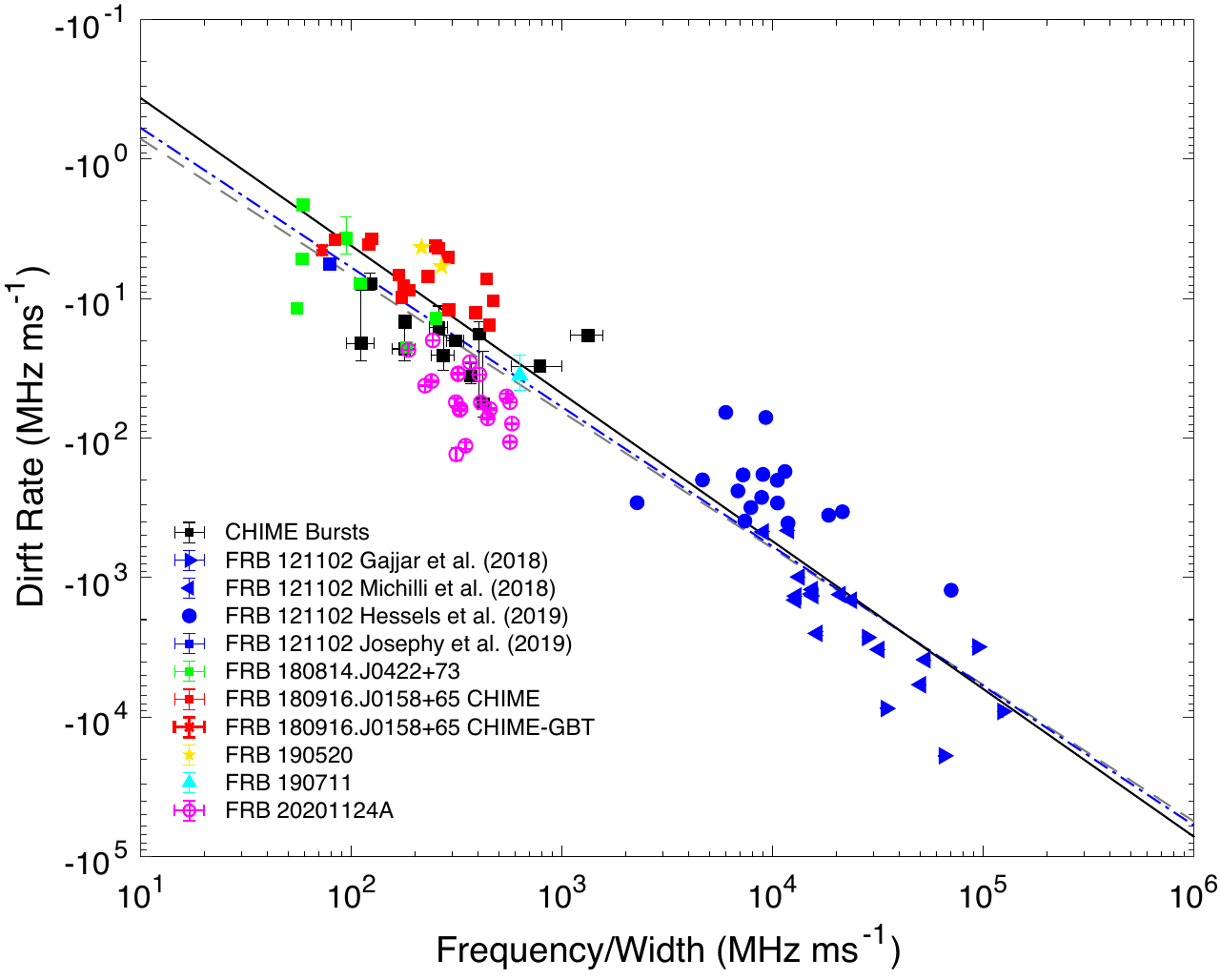}
\caption{\small{Same as Figure \ref{fig:ratevsnu} but for drift rates at different ratio of frequency to sub-pulse width.}}
\label{fig:ratevswidth}
\end{figure}

As the magnetosphere rotates, magnetic field lines sweep
across the LOS and there may be more than one bulk of bunches being observed coincidentally.
As shown in Figure \ref{fig:sub-pulse}, line 1 and line 2 are approximately in the same plane due to slow rotation.
However, FRB sources can repeat so that it is hard to distinguish whether the observed two pulses are the sub-pulses of one burst or indeed two individual pulses.
The ``sub-pulse'' emission discussed here, is intrinsically identified as bursts emitted by sparking events from the same trigger event on the stellar surface.
The trigger event may last a long time, but sub-pulses are observed within one period.

The temporal and frequency properties of sub-pulses are
supposed to be strongly related to the magnetospheric geometry.
The time delay for the two waves reads
\be
\Delta t=t_{20}-t_{10}+\Delta t_{\mathrm{geo}},
\label{eq:deltat}
\ee
where $t_{10}$ and $t_{20}$ are the times when the bunches are generated, and $\Delta t_{\mathrm{geo}}$ is the geometric time delay, which can be calculated as
\be
\Delta t_{\mathrm{geo}}=\frac{s_{2}-s_{1}}{v_{e,\parallel}}-\frac{\Delta r_p \cos \theta_{p}}{c},
\label{eq:tgeo}
\ee
where $\Delta r_p$ denotes the distance between the points where emission can sweep the LOS at two magnetic field lines, i.e. two emitting points, and $v_{e,\parallel}$ is the velocity of the positrons parallel to the B-field lines \citep{Wang20}.
From Equation (\ref{eq:Theta}), the distance $\Delta r_p=\Delta r$ if the neutron star rotates very slowly, i.e., $\Delta t_{\rm obs}\ll P$.
The velocity $v_{e,\parallel}\simeq\beta_e c$ when charges are far from the light cylinder.
Footpoints of the field lines are located in the polar region, so that we have $\theta_0\ll 1$ and then Equation (\ref{eq:tgeo}) can be written as
\be
\Delta t_{\rm geo}\simeq\frac{\Delta r}{c}\left[\left(1+\frac{1}{2\gamma^2}\right)I_n(\theta)-\cos\theta_p\right],
\label{eq:tgeo2}
\ee
where $I_n(\theta)$ is a dimensionless parameter denoting the enhancement factor due to the field line curvature (see Appendix \ref{app:geometry}).

First, we consider that charges are triggered at the same time, i.e., $t_{20}=t_{10}$.
Combining Equation (\ref{eq:Ebalance}), (\ref{eq:tgeo2}) and the specific geometric condition of magnetic field, one can obtain
\be
{\dot \nu}=\left[\frac{1}{2}\frac{\Delta\rho}{\rho\Delta t}+\frac{3}{4}\left(\frac{\partial E_{\|}}{E_{\|}\partial r}\frac{\Delta r}{\Delta t}+\frac{\partial E_{\|}}{E_{\|}\partial t}\right)\right]\nu.
\label{eq:driftrate}
\ee
In the following discussion, we assume that $E_{\|}$ is independent with time, i.e., $\partial E_{\|}/\partial t=0$.
From Equation (\ref{eq:Ebalance}), in order to derive a constant Lorentz factor, one requires that $E_{\|}\propto r^{-2}$.
Then, the drift pattern can be described as $\dot{\nu}\propto -\nu^2$.
More complicated time-frequency structure could originate from $E_{\|}$ with more complex formulae.

A bunch roughly has a thickness of half wavelength, contributing to the FRB radiation for a time duration of the order of $\nu^{-1}$ in the observer frame.
The thickness becomes larger as the bunch moves to a higher altitude.
According to Equation (\ref{eq:duration}), if the neutron star rotates slowly, a continuous plasma flow emits for a duration of $w\propto N_B\nu^{-1}$, where $N_B$ is the total number of bunch generated during the trigger.
The pulse width is related to $\nu$, e.g., $w\propto \nu^{-1}$, which leads to $\dot{\nu}\propto -\nu/w$.

We investigate the relationship between drift rate and burst central frequency.
All parameters are considered in the rest frame of the host galaxy.
Therefore, one has $\nu=\nu_{\rm obs}(1+z)$, $\dot{\nu}=\dot{\nu}_{\rm obs}(1+z)^2$ and $w=w_{\rm obs}/(1+z)$.
The observed width is derived from autocorrelation functions by fitting a two-dimensional Gaussian \citep{Chamma21}.
Most repeaters have measured redshifts.
For the sources without precise redshift measurements, their redshifts are calculated from the dispersion measure (DM) models (see, e.g., \citealt{Deng14,Zhang18,Pol19,Cordes21}).

The relationship between $\dot{\nu}$ and central frequency as well as $\dot{\nu}$ and the ratio of $\nu/w$, are plotted in Figure \ref{fig:ratevsnu} and \ref{fig:ratevswidth}.
These relationships have been used to fit the data for several repeaters \citep{Chamma21}.
Here, we expand our sample and apply these observed results in the rest frame of their host galaxies.
The data are quoted from \cite{Gajjar18,Michilli18,CHIME19a,CHIME19b,Hessels19,Josephy19,Caleb20,Chamma21,Chawla20,Day20,Pastor20,Hilmarsson21b,Platts21}.
Note that the observed central frequencies for the CHIME bursts (black squares in Figure \ref{fig:ratevsnu} and \ref{fig:ratevswidth}) are adopted as 600 MHz and 1370 MHz for FRB 201124A \citep{Hilmarsson21b}.
The details of data analysis for FRB 190520 are shown in Appendix \ref{app:190520}.

We fit both the $\dot{\nu}-\nu$ and $\dot{\nu}-\nu/w$ relationships.
The fitting is firstly considered for FRB 121102, which exhibits the most multi-band data.
Both the $\dot{\nu}-\nu$ and $\dot{\nu}-\nu/w$ relationships can be well fitted by power laws:
$\dot{\nu}=-5.0^{+83.1}_{-4.7}\times10^{-6}\nu^{2.29\pm0.36}$ and $\dot{\nu}=-3.2^{+57.0}_{-3.0}\times10^{-2}(\nu/w)^{1.06\pm0.31}$.
We then fix the power law index as 2 and 1 for $\dot{\nu}-\nu$ and $\dot{\nu}-\nu/w$ fitting, and the results are given by $\dot{\nu}=-5.4^{+1.4}_{-1.1}\times10^{-5}\nu^2$ and $\dot{\nu}=-5.5^{+2.5}_{-1.7}\times10^{-2}\nu/w$.
Finally, we fit both relationships to power laws for all FRBs.
The results are the following:
$\dot{\nu}=-1.1^{+3.4}_{-0.8}\times10^{-6}\nu^{2.45\pm0.21}$ and
$\dot{\nu}=-7.6^{+7.3}_{-3.7}\times10^{-2}(\nu/w)^{0.97\pm0.09}$.

\subsection{Subpulse interval}\label{sec3.2}

\begin{figure}
\includegraphics[width=0.48\textwidth]{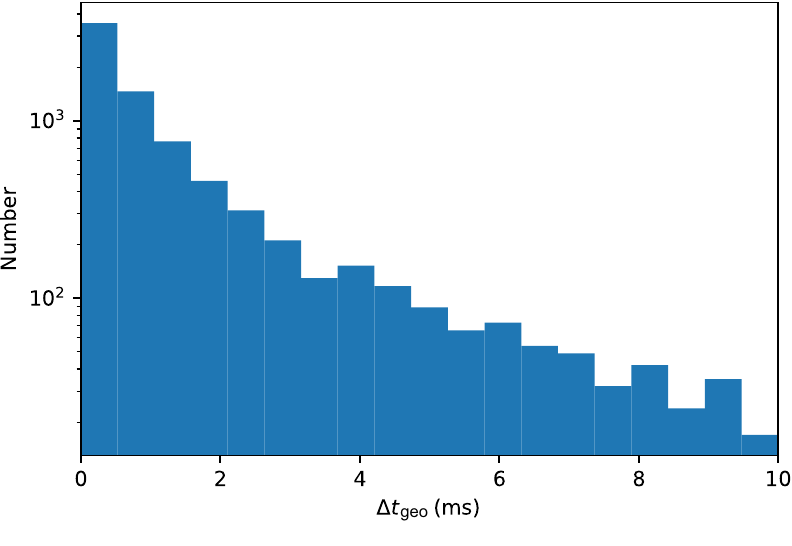}
\caption{\small{Histograms of estimated $\Delta t_{\rm obs}$ for quadrupole from Equation (\ref{eq:tdistribution}).}}
\label{fig:deltat}
\end{figure}

There are rare FRBs exhibiting upwards-drifting structures.
Regardless of whether two or more pulses are sub-pulses of a single burst or truly multiple bursts, the observed interval is generally described by Equation (\ref{eq:deltat}).
Consequently, if the trigger event has a long-duration, there would be a higher chance to observe upward-drifting events \citep{Wang20}.
For a short-duration triggering mechanism ($|t_{20}-t_{10}|\ll\Delta t_{\rm geo}$), the observed interval for two sub-pulses is determined by
\be
\Delta t_{\rm geo}=I_n(\theta)\frac{R}{nc}\left[\frac{f(\theta)}{f(\theta_s)}\right]^{1/n}\Delta {\rm ln}[f (\theta_s)].
\label{eq:tdistribution}
\ee
Based on Equation (\ref{eq:tdistribution}), we plot the simulated distribution of the observed time intervals
for a quadrupole field with $10^4$ samples, as shown in Figure \ref{fig:deltat}.
We fix $\theta$ at $\theta_{\rm max}$ for the quadrupole.
The events which $\Delta t_{\rm geo}<0.1\,\rm ms$ are ignored.
Two angles $\theta_{s1}$ and $\theta_{s2}$ are
uniformly distributed in the range of $0-0.06$\footnote{The maximum $\theta_s=0.06$ is adopted, which is $\theta_{\rm max}$ at the location of $r=10R$.} rad in the simulations.
The simulated interval is generally consistent with the sub-burst separation distribution from \cite{Pleunis21}.

\subsection{Radius-to-frequency mapping}\label{sec3.3}

\begin{figure*}
\begin{center}
\includegraphics[width=0.98\textwidth]{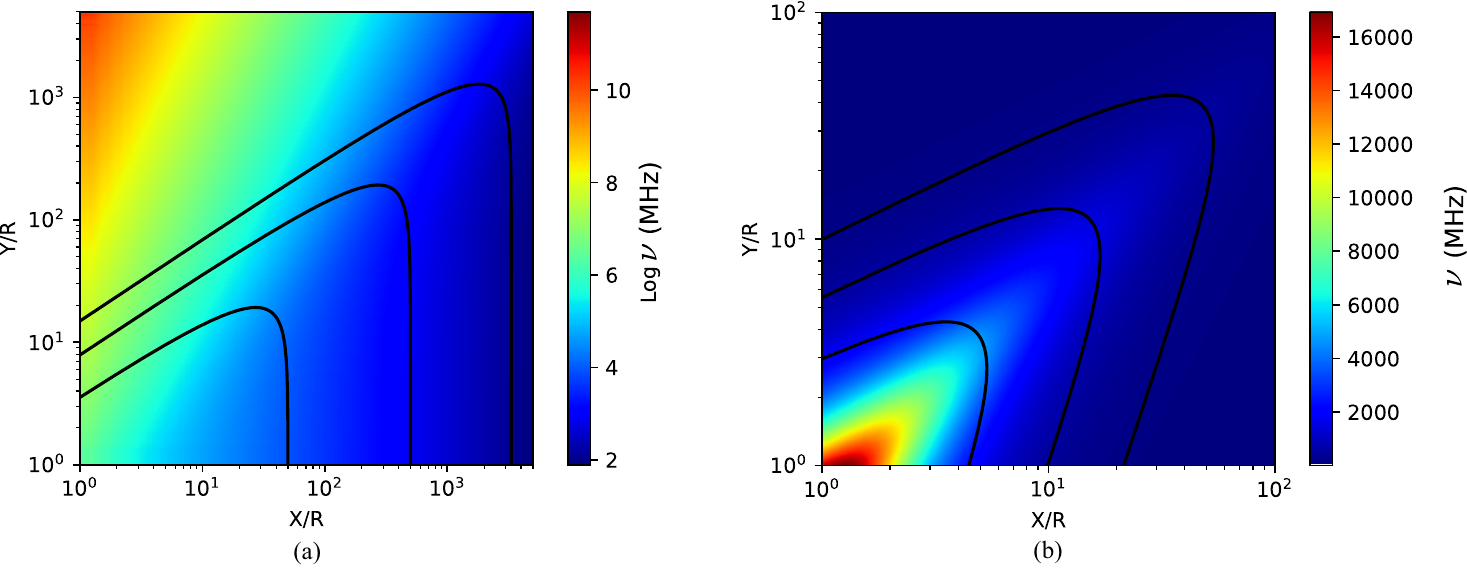}
\caption{\small{Radius-to-frequency mapping derived by the drifting pattern for two  cases: (a) dipole; (b) quadrupole.
Black solid lines denote magnetic field lines.}}
\label{fig:mapping}
\end{center}
\end{figure*}

The drifting pattern of FRBs reflects the radiation frequency change with the location in the magnetosphere, i.e. the so-called ``radius-to-frequency mapping'' \citep[e.g.][]{Wang19}.
For all sparking particles created at the same time, if $\gamma$ is a constant, combining  Equation (\ref{eq:tgeo}), (\ref{eq:driftrate}) and specific geometry condition as discussed in Section \ref{sec2.1}, one can derive
\be
\dot{\nu}=-\frac{4\pi C_n(\theta)}{3\gamma^3[I_n(\theta)-\cos\theta_p]}\nu^2,
\label{eq:mapping}
\ee
where $C_n(\theta)$ is a geometric factor (see Equation (\ref{eq:curvatureradius0}) in Appendix).

We first consider a dipolar configuration.
Combining the fitting results of FRB 121102 with Equation (\ref{eq:mapping}), we plot the radius-to-frequency mapping of the magnetosphere, shown in panel (a) of Figure \ref{fig:mapping}.
The observed FRB frequency ranges from $100$ MHz to $\gtrsim8$ GHz.
The pulse-to-pulse temporal property is caused by the geometric effect, i.e., different emitting bunches (sparks) traveling in different trajectories exhibit different photon arrival delays.
However, aberrations of light can reduce the trajectory-induced interval.
From Equation (\ref{eq:tgeo}), we can obtain $\Delta r\simeq c\Delta t_{\rm geo}\sin^2\theta$ for $\theta\sim0.1$.
In order to create millisecond burst intervals, the distance between two emission regions should be $\Delta r\sim10^3R$.

Another possible case is that radio waves are produced at $\theta\simeq\pi/2$.
The aberration term could be neglected for a dipolar configuration when $\theta\sim\pi/2$, so that we have $\Delta r_{\rm geo}\simeq c\Delta t_{\rm obs}/I_1(\pi/2)$.
The emission region is located at $r\gtrsim10 R$.

We then consider a quadrupolar configuration.
Based on Equation (\ref{eq:mapping}), the spectro-temporal properties strongly depends on the polar angle of LOS.
However, which trajectory plane truly sweeps across the LOS is random.
If $\theta>\theta_{\rm max}$, particles would move toward the stellar surface, resulting in the radiation beam not swept by the LOS.
For $\theta\ll0.5$, the quadrupole has similar geometric properties as the dipole.
Beyond that region, quadruplar field lines are more curved, so that the particles should travel a longer distance than the dipole case to reach the emitting points.
For $\pi/5\lesssim\theta$, the aberration term is not an order of magnitude larger than the interval traveled.
The interval distance is $\Delta r\simeq c\Delta t_{\rm obs}/I_2(\theta_{\rm max})$.
In order to generally match the fitting results, we adopt $\gamma=100$ and plot the radius-to-frequency mapping, shown in panel (b) of Figure \ref{fig:mapping}.
FRBs are suggested to be emitted at several to hundreds of the stellar radius.

\section{Conclusion and Discussion}\label{sec4}

In this paper, we developed a general radiation theory of coherent curvature radiation of charged bunches in the magnetosphere from the first principle, and applied the model to interpret the spectral and polarized emission properties of repeating FRBs.
Within this framework, we consider the spectro-temporal pattern by deriving a generic magnetic configuration, and modeled the observed drift rates for some FRBs.
The following conclusions can be drawn: 

(1) We consider a general scenario that charged bunches move in a rotating magnetosphere with a multipolar magnetic field.
The emitting bunches has a specific configuration as defined by the magnetic field lines.
Trajectories (field lines) for charges are not
parallel to each other, which allows the angle of LOS with respect to the central axis entering the problem.
Combining with the magnetic field configuration
considered, we calculated the predicted radiation spectra, which can be generally characterized by multi-segment broken power laws, and evolve with the motion of the and the rotation of the neutron star.
The total flux for an on-beam case is higher than that of an off-beam case.

(2) For uniformly distributed particle in bunches, the emission is 100\% linearly polarized when the LOS is parallel to the central axis.
The sign of circular polarization will change when the LOS is parallel to the symmetric axis of the bulk.
If the opening angle of the bulk is smaller than $1/\gamma$ ($\omega_c<\omega_t$), that is, on-beam case, the emission keeps being highly linearly polarized as the LOS sweeps across the emitting bunches, with a flat PA evolution.
The circular polarization degree becomes stronger as $|\varphi|$ grows, due to the non-axisymmetric summation of $A_{\perp}$.
High circular polarization events as rarely observed suggest that most bunches have small open angles in those cases.
PA evolution is more dramatic than that for the on-beam case.
A luminous and highly circular polarized pulse may be generated when $\varphi_t\sim1/\gamma$ or when particles are extremely non-uniformly distributed in bunches.
Both linear and circular polarization degrees are frequency dependent and the waves are 100\% polarized waves.

(3) The radiation front pushes aside the pair plasma and can break out of the magnetosphere, forming a charge starvation region.
In this region, an electric field might be produced and quickly balanced with the electric power provided by $E_{\|}$.
The required $E_{\|}$ is smaller than that proposed by \cite{Kumar17}, where only the kinetic energy was considered.

(4) We expand the geometric model \citep{Wang19,Wang20}, where only the dipolar field was considered, to a more general scenario by invoking multi-polar fields.
Within the framework of curvature radiation, the spectro-temporal property of sub-pulses can be naturally modeled by the geometric model.
Such drifting structure would appear at some of single pulses from a pulsar with FRB-like trigger mechanism.
For a constant $\gamma$, regardless of the magnetic configuration, the observed drift rate generally follows $\dot{\nu}\propto-\nu^2$.
The drift rate as a function of frequency for FRB 121102 can be well fitted by the model and a generic power law with index $2.29\pm0.36$.
For the sample of all FRBs, the best power law fitting result is $2.45\pm0.19$.
The drift rate as a function of $\nu/w$ can be also fitted by a power law.
The indices are $1.06\pm0.31$ for FRB 121102  and $0.97\pm0.09$ for all FRBs.

(5) Based on the geometric model, the spectro-temporal properties suggest that the radiation site varies in the magnetosphere.
We derive the radius-to-frequency mapping within both dipolar and quadrupolar scenarios.
For the dipolar scenario, the observed FRBs are suggested to be emitted at $r\gtrsim10R$ in closed field lines.
This is a just result by satisfying geometric conditions.
A quadrupolar configuration has more curved field lines, so that emission height is at several to hundreds stellar radii, lower than that of the dipolar scenario.
FRB sources are most likely to host multipolar fields or more curved magnetic configurations than an ideal dipole.

According to the rotation vector model \citep{Radhakrishnan69}, the PA is generally flat for a slow-rotating pulsar, but evolves at near the smallest impact angle.
Within this scenario, the flatness of PA evolution and extremely high linear polarization degree of the emission of FRBs 121102, 180916 and 20201124A suggest that neutron stars rotate slowly ($P>1$ s) with a small opening angle.
Flat PA evolution but with $\sim10\%$ circular polarization when $\varphi_t=0.1\gamma^{-1}$ match the observation of FRB 20201124A.
If the multipolar components have the same axis as the dipole, the PA evolution can also be characterized by the ``S-shape'' rotation vector model.
Variable PA evolution (FRB 180301, 181112) may require more complicated magnetic field configurations and LOS geometry, or  propagation effects operating either inside the magnetosphere or far from the source (e.g., \citealt{Dai20}).

A triple sub-pulse burst has been found in FRB 121102 (burst 6, MJD 58075, see \citealt{Hilmarsson21a}).
An apparent upward drifting pattern can be seen between the first two components, while the second and third components exhibit a downward drifting pattern.
The PA of the first component differs from those of the other two.
Based on the geometric model, as the magnetosphere sweeps across the LOS, one would first observe an upward drifting pattern, and after the LOS crosses the minimum impact angle, a downward drifting pattern.
During this process, PA evolves as ``S-shape'' according to the classical rotation vector model.

According to Equation (\ref{eq:driftrate}), the fluctuation of the drift rate may be related to the fluctuation of the particle number and the inclination of the LOS.
Most repeating FRBs have the same order of drift rate at same central frequency, suggesting that these
sources most likely have similar magnetospheres.
Surprisingly, such drift pattern has also been discovered in at least some single pulses of PSR B0950+08, but with lower frequency bands (20--83 MHz) \citep{Bilous21}.
\cite{Rajabi20} proposed a dynamical model, in which drift rate is a function of $\nu/w$ because of the relativistic motion.
However, within the curvature radiation model, the burst duration depends on the number that bunches travel trough during the LOS sweeps the radiation region, e.g., $w\sim N_B\nu^{-1}$.
Consequently, the relationship $\dot{\nu}\propto\nu/w$ is given rise to $\dot{\nu}\propto\nu^2$.

This paper focuses on coherent curvature radiation by bunches. Recently, \cite{Zhang21} proposed another general classes of models invoking coherent inverse Compton scattering by bunches. The spectra and polarization properties of that model require detailed investigations.

\section*{Acknowledgments}
The authors are grateful to Mohammed Chamma and Henning Hilmarsson for providing data.
We are also grateful to Xuelei Chen, Shi Dai, Clancy James, Jinchen Jiang, Jiguang Lu, Rui Luo, Donald Melrose, Lijing Shao, Jumpei Takata, Hao Tong, Zhen Yan, Jumei Yao, Haoyang Ye, Bin-Bin Zhang, Song-Bo Zhang and an anonymous referee for helpful comments and discussions.
This work is supported by the National Key R\&D program of China No. 2017YFA0402602 and National SKA Program of China No. 2020SKA0120100.
W.-Y.W. is supported by a Boya Fellowship and the fellowship of China Postdoctoral Science Foundation No. 2021M700247.
Y.-P.Y. is supported by NSFC grant No. 12003028.
C.-H.N. is supported by a FAST Fellowship.
R.X. is supported by strategic Priority Research Program of CAS (XDB23010200).

\appendix

\section{Magnetosphere Geometry}\label{app:geometry}

In this section, we briefly summarize the geometric properties of magnetic configuration.
Generally, with the assumption of axisymmetric force-free condition, the flux function obeys the Grad-Shafranov equation
\begin{equation}
\frac{\partial^{2} A}{\partial r^{2}}+\frac{1-x^{2}}{r^{2}} \frac{\partial^{2} A}{\partial x^{2}}+F(A) \frac{d F}{d A}=0,
\label{eq:Grad-Shafranov}
\end{equation}
where $x=\cos\theta$ and $F(A)$ is the free function.
The magnetic field follows from $F(A)$ and $A$:
\begin{equation}
\boldsymbol{B}=\frac{1}{r \sin \theta}\left[\frac{1}{r} \frac{\partial A}{\partial \theta} \hat{\boldsymbol{r}}-\frac{\partial A}{\partial r} \hat{\boldsymbol{\theta}}+F(A) \hat{\boldsymbol{\phi}}\right].
\end{equation}
For an untwisted magnetosphere, $F(A)=0$.
With the separable solutions of the form
\begin{equation}
A=r^{-n} f(\cos\theta),
\end{equation}
there is an ordinary differential equation \citep{Low90}
\begin{equation}
\left(1-x^{2}\right) f^{\prime \prime}(x)+n(n+1) f(x)=0,
\end{equation}
where $n$ is a constant denoting the field configuration, i.e., $n=1$ for dipole, $n=2$ for quadrupole etc.
Therefore, the magnetic field can be described as
\be
\boldsymbol{B}=\frac{1}{r^{2+n}}\left[-\frac{\partial f}{\partial \cos\theta} \hat{\boldsymbol{r}}+n\frac{f}{\sin\theta} \hat{\boldsymbol{\theta}}\right].
\label{eq:B-field}
\ee
The magnetic field lines satisfy
\be
\frac{dr}{rd\theta}=\frac{\partial f}{n f\partial \theta}.
\ee
One can define $R_{\rm max}$, which denotes the largest distance from the stellar center at which the field line crosses the plane with colatitude of $\theta_{\rm max}$.
For certain field line with $R_{\rm max}$, we can obtain
\be
\frac{r}{R_{\rm max}}=\left[\frac{f(\theta)}{f(\theta_{\rm max})}\right]^{1/n},
\label{eq:Bgeo}
\ee
and
\be
\tan\Theta=\frac{f'\sin\theta+nf\cos\theta}{f'\cos\theta-nf\sin\theta}.
\label{eq:Theta}
\ee
where $\Theta$ is the angle between the magnetic axis and the tangent direction of magnetic field lines, and $f'$ denotes the derivative of $f$ with respect to $\theta$.
From Equation (\ref{eq:Theta}), the tangential direction for a certain field line only depends on $\theta$ rather than $r$.
Basically, the curvature radius at $(r,\,\theta)$ is given by
\be
\rho=\frac{\left(r^{2}+r^{\prime 2}\right)^{3 / 2}}{\left|r^{2}+2 r^{\prime 2}-r r^{\prime \prime}\right|}=C_n(\theta)r,
\label{eq:curvatureradius}
\ee
where
\be
C_n(\theta)=\frac{[1+(f'/nf)^2]^{3/2}}{|1+(f'/nf)^2-f''/nf+f'^2/nf^2|}.
\label{eq:curvatureradius0}
\ee
The distance that charges travel from $\theta_0$ to $\theta$ is given by
\be
s=\int^\theta_{\theta_0}\sqrt{dr^2+r^2d\theta^2}=rI_n(\theta,\theta_0),
\label{eq:distance}
\ee
where 
\be
I_n(\theta,\theta_0)=\left[\frac{1}{f(\theta)}\right]^{1/n}\int^\theta_{\theta_0}\left[f(\theta')\right]^{1/n}\sqrt{1+\frac{1}{n^2f(\theta')^2}\left(\frac{\partial f}{\partial\theta'}\right)^2}d\theta',
\label{eq:Itheta}
\ee
is a dimensionless parameter denoting the enhancement factor due to field line curvature.
We define $I(\theta)=I_n(\theta,0)$.
The angle between radial direction and tangent direction of magnetic field lines is given by
\be
\cos\theta_p=\frac{f'}{\sqrt{f'^2+n^2f^2}}.
\label{eq:thetap}
\ee
In general, a bunch shown in Figure \ref{fig:charges}, can be simply adopted as a cuboid.
Their length, height and width are approximately $\lambda$, $r\Delta\theta$ and $2\varphi_tr$, leading to the cuboid volume of $V_b\simeq2\varphi_tr^2\lambda\Delta\theta\simeq2\varphi_t r^2 \lambda[f'(\theta_s)/f'(\theta)]\Delta \theta_s$.

\section{Coherent Radiation from Moving Charges}\label{app:coherent}

\begin{figure*}
\begin{center}
\includegraphics[width=0.5\textwidth]{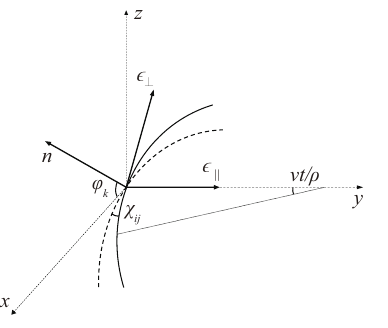}
\caption{\small{Geometry for instantaneous circular motion. The trajectory lies in the $x$-$y$ plane. At the retarded time $t=0$, the positron is at the origin. The dashed trajectory shows a positron at the origin, and the velocity
is along the $x$-axis. The angle between the solid and dashed trajectory is $\chi_{ij}$ at $t=0$. The angle between LOS and the trajectory plane is $\varphi_k$.
}}
\label{fig:xyz}
\end{center}
\end{figure*}

In this section, we briefly summarize the physics of radiation from moving charges.
Define $\boldsymbol{\beta}_{e\perp}$ as the component of $\boldsymbol{\beta}_e$ in the plane that is perpendicular to the LOS:
\begin{equation}
\boldsymbol{\beta}_{e\perp}=-\boldsymbol{n} \times\left(\boldsymbol{n} \times \boldsymbol{\beta}_{e}\right).
\end{equation}
For a single charge, the energy radiated per unit solid angle per unit frequency interval is given by (e.g., \citealt{Rybicki79,Jackson98})
\be
\frac{d^2W}{d \omega d \Omega}=\frac{e^{2} \omega^{2}}{4 \pi^{2} c}\left|\int_{-\infty}^{+\infty} -\boldsymbol{\beta}_{e\perp}  e^{i \omega(t-\boldsymbol{n} \cdot \boldsymbol{r}(t) / c)} d t\right|^{2}=\frac{e^{2} \omega^{2}}{4 \pi^{2} c}\left|-\boldsymbol{\epsilon}_{\|} A_{\|}+\boldsymbol{\epsilon}_{\perp} A_{\perp}\right|^{2},
\label{eq:dI/domegadOmega}
\ee
where $\boldsymbol{\epsilon}_{\|}$ is the unit vector pointing to the center of the instantaneous circle, $\boldsymbol{\epsilon}_{\perp}=\boldsymbol{n}\times\boldsymbol{\epsilon}_{\|}$ is defined, and $A_{\|}$ and $A_{\perp}$ are the polarized components of the amplitude along $\boldsymbol{\epsilon}_{\|}$ and $\boldsymbol{\epsilon}_{\perp}$, respectively.

As shown in Figure \ref{fig:xyz}, for one trajectory, the angle between the electron velocity direction and $x$-axis at $t=0$ is defined as $\chi_{ij}$, and that between the LOS and the trajectory plane is $\varphi_k$.
Therefore, we have
\be
\boldsymbol{n} \times\left(\boldsymbol{n} \times \boldsymbol{\beta}_{e, ijk}\right)=\beta_e\left[-\boldsymbol{\epsilon}_{\|} \sin \left(\frac{v_e t}{\rho}+\chi_{ij}\right)+\boldsymbol{\epsilon}_{\perp} \cos \left(\frac{v_e t}{\rho}+\chi_{ij}\right) \sin \varphi_k\right],
\ee
and
\begin{equation}
\begin{array}{c}
\omega\left(t-\frac{\boldsymbol{n} \cdot \boldsymbol{r}_{ijk}(t)}{c}\right)
=\omega\left[t-\frac{2 \rho}{c} \sin \left(\frac{v_e t}{2 \rho}\right) \cos \left(\frac{v_e t}{2 \rho}+\chi_{ij}\right) \cos \varphi_k\right]\\
\simeq \frac{\omega}{2}\left[\left(\frac{1}{\gamma^{2}}+\varphi_k^{2}+\chi_{ij}^{2}\right) t+\frac{c^{2} t^{3}}{3 \rho^{2}}+\frac{c t^{2}}{\rho} \chi_{ij}\right].
\end{array}
\end{equation}
The amplitudes for one positron are given by
\be
\begin{aligned}
&A_{\|, ijk} \simeq \int_{-\infty}^{\infty}\left(\frac{c t}{\rho}+\chi_{ij}\right) \exp \left(i \frac{\omega}{2}\left[\left(\frac{1}{\gamma^{2}}+\varphi_k^{2}+\chi_{ij}^{2}\right) t+\frac{c^{2} t^{3}}{3 \rho^{2}}+\frac{c t^{2}}{\rho} \chi_{ij}\right]\right) d t, \\
&A_{\perp, ijk} \simeq \varphi_k \int_{-\infty}^{\infty} \exp \left(i \frac{\omega}{2}\left[\left(\frac{1}{\gamma^{2}}+\varphi_k^{2}+\chi_{ij}^{2}\right) t+\frac{c^{2} t^{3}}{3 \rho^{2}}+\frac{c t^{2}}{\rho} \chi_{ij}\right]\right) d t .
\end{aligned}
\ee
The introduction of the parameter $u$ and $\xi$, i.e.
\begin{equation}
\begin{aligned}
&u=\frac{c t}{\rho}\left(\frac{1}{\gamma^{2}}+\varphi_{k}^{2}+\chi_{ij}^2\right)^{-1 / 2}, \\
&\xi=\frac{\omega \rho}{3 c}\left(\frac{1}{\gamma^{2}}+\varphi_{k}^{2}+\chi_{ij}^2\right)^{3 / 2},
\end{aligned}
\end{equation}
allows us to transform the the polarized components of the amplitude into the form:
\begin{equation}
\begin{aligned}
&A_{\|, ijk} \simeq \frac{\rho}{c}\left(\frac{1}{\gamma^{2}}+\varphi_{k}^2+\chi_{ij}^2\right) \int_{-\infty}^{\infty}\left(u+\frac{\chi_{ij}}{\sqrt{1 / \gamma^{2}+\chi_{ij}^{2}+\varphi_k^2}}\right) \exp \left(i \frac{3}{2} \xi\left(u+\frac{1}{3} u^{3}+\frac{\chi_{ij}}{\sqrt{1 / \gamma^{2}+\varphi_k^2+\chi_{ij}^{2}}} u^{2}\right)\right) d u,\\
&A_{\perp, ijk} \simeq
\frac{\rho}{c}\varphi_k\left(\frac{1}{\gamma^{2}}+\varphi_{k}^2+\chi_{ij}^2\right)^{1/2} \int_{-\infty}^{\infty}\exp \left(i \frac{3}{2} \xi\left(u+\frac{1}{3} u^{3}+\frac{\chi_{ij}}{\sqrt{1 / \gamma^{2}+\varphi_k^2+\chi_{ij}^{2}}} u^{2}\right)\right) d u.\\
\end{aligned}
\end{equation}
Note that
\be
\begin{aligned}
&\lim\limits_{u\to+0} \left(u+\frac{1}{3} u^{3}+\frac{\chi_{ij}}{\sqrt{1 / \gamma^{2}+\varphi_k^2+\chi_{ij}^{2}}} u^{2}\right)=u,\\
&\lim\limits_{u\to+\infty} \left(u+\frac{1}{3} u^{3}+\frac{\chi_{ij}}{\sqrt{1 / \gamma^{2}+\varphi_k^2+\chi_{ij}^{2}}} u^{2}\right)=\frac{1}{3}u^3.
\end{aligned}
\ee
Therefore, the amplitudes in the above equation are given by
\begin{equation}
\begin{aligned}
&A_{\|, ijk} \simeq \frac{i2}{\sqrt{3}}\frac{\rho}{c}\left(\frac{1}{\gamma^{2}}+\varphi_{k}^2+\chi_{ij}^2\right)K_{\frac{2}{3}}(\xi)+ \frac{2}{\sqrt{3}}\frac{\rho}{c}\chi_{ij}\left(\frac{1}{\gamma^{2}}+\varphi_{k}^2+\chi_{ij}^2\right)^{1/2}K_{\frac{1}{3}}(\xi),\\
&A_{\perp, ijk} \simeq \frac{2}{\sqrt{3}}\frac{\rho}{c}\varphi_k\left(\frac{1}{\gamma^{2}}+\varphi_{k}^2+\chi_{ij}^2\right)^{1/2}K_{\frac{1}{3}}(\xi),
\end{aligned}
\end{equation}
where $K_\nu(\xi)$ is the modified Bessel function.

For a single charge with identifiers  $i,j,k$, the energy radiated per unit solid angle per unit frequency interval can be written as
\begin{equation}
\frac{d^2W}{d \omega d \Omega}\simeq \frac{e^{2}}{3 \pi^{2} c}\left(\frac{\omega \rho}{c}\right)^{2}\left(\frac{1}{\gamma^{2}}+\varphi_{k}^2+\chi_{ij}^2\right)^{2}\left[K_{2 / 3}^{2}(\xi)+\frac{\varphi_k^{2}+\chi_{ij}^2}{1/\gamma^{2}+\varphi_{k}^2+\chi_{ij}^2} K_{1 / 3}^{2}(\xi)\right].
\end{equation}
The circular polarization degree is given by
\be
\frac{V}{I}=\frac{2\varphi_k\left(1/\gamma^{2}+\varphi_{k}^2+\chi_{ij}^2\right)^{1/2}K_{2 / 3}(\xi)K_{1 / 3}(\xi)}{K_{2 / 3}^{2}(\xi)/\gamma^2+(\varphi_k^2+\chi_{ij}^2)[K_{2/ 3}^{2}(\xi)+K_{1/ 3}^{2}(\xi)]}.
\ee
The 100\% linear polarization appears at $\varphi_k=0$ and circular polarization becomes stronger as $\varphi_k$ gets larger.

We assume that the curvature radius is a constant in the bulk and each bunch has roughly the same $\chi_{u,i}$ and $\chi_{d,i}$.
Then, one can derive average heights to replace the complex boundary condition of the bulk, shown as Figure \ref{fig:charges}.
For more than one charged particle, a coherent summation of
the amplitudes should replace the single amplitude.
The total amplitude of $i$th bunch projection at the plane perpendicular to the LOS is then given by
\be
\begin{aligned}
&A_{\|, i}\simeq\frac{2}{\sqrt{3}}\frac{\rho}{c}\frac{N_\theta}{\Delta\theta_s}\frac{N_\phi}{2\varphi_t}\int^{\chi_{u,i}}_{\chi_{d,i}}d\chi'\int^{\varphi_u}_{\varphi_d}\left[i\left(\frac{1}{\gamma^{2}}+\varphi'^2+\chi'^2\right)K_{\frac{2}{3}}(\xi)+ \chi'\left(\frac{1}{\gamma^{2}}+\varphi'^2+\chi'^2\right)^{1/2}K_{\frac{1}{3}}(\xi)\right]\cos\varphi'd\varphi',\\
&A_{\perp, i} \simeq \frac{2}{\sqrt{3}}\frac{\rho}{c}\frac{N_\theta}{\Delta\theta_s}\frac{N_\phi}{2\varphi_t}\int^{\chi_{u,i}}_{\chi_{d,i}}d\chi'\int^{\varphi_u}_{\varphi_d}\left(\frac{1}{\gamma^{2}}+\varphi'^2+\chi'^2\right)^{1/2}K_{\frac{1}{3}}(\xi)\varphi'\cos\varphi'd\varphi'.
\end{aligned}
\ee

For any $\varphi'\ll1/\gamma$ and $\chi'\ll1/\gamma$, the radiation emits within a conal angle $1/\gamma$.
If $\omega\gg\omega_c$, one has $\xi\approx\omega/(2\omega_c)\gg1$, leading to $K_{\nu}(\xi)\rightarrow \sqrt{\pi / 2 \xi} \exp (-\xi)$. Therefore, one has approximately
\begin{equation}
\begin{aligned}
&A_{\|,i}\simeq(\varphi_u-\varphi_d)\frac{N_\theta}{\Delta\theta_s}\frac{N_\phi}{2\varphi_t}\frac{2}{\sqrt{3}} \frac{\rho}{\gamma c}\sqrt{\frac{\pi\omega_c}{\omega}}\exp\left(-\frac{\omega}{2\omega_c}\right)\left[i\frac{\chi_{u,i}-\chi_{d,i}}{\gamma}+\frac{\chi_{u,i}^2-\chi_{d,i}^2}{2}\right], \\
&A_{\perp,i}\simeq\frac{N_\theta}{\Delta\theta_s}\frac{N_\phi}{2\varphi_t}\frac{2}{\sqrt{3}} \frac{\rho }{\gamma c}\sqrt{\frac{\pi\omega_c}{\omega}}\exp\left(-\frac{\omega}{2\omega_c}\right)(\chi_{u,i}-\chi_{d,i})\left[\frac{\varphi_u^2-\varphi_d^2}{2}\right].
\end{aligned}
\end{equation}
If $\omega\ll\omega_c$, the amplitudes read
\be
\begin{aligned}
&A_{\|,i}\simeq(\varphi_u-\varphi_d)\frac{N_\theta}{\Delta\theta_s}\frac{N_\phi}{2\varphi_t}\frac{1}{\sqrt{3}} \frac{\rho}{\gamma c}\left[i\frac{\chi_{u,i}-\chi_{d,i}}{\gamma}\Gamma(2/3)\left(\frac{\omega}{4\omega_c}\right)^{-2/3}+\frac{\chi_{u,i}^2-\chi_{d,i}^2}{2}\Gamma(1/3)\left(\frac{\omega}{4\omega_c}\right)^{-1/3}\right], \\
&A_{\perp,i}\simeq\frac{N_\theta}{\Delta\theta_s}\frac{N_\phi}{2\varphi_t}\frac{1}{\sqrt{3}} \frac{\rho }{\gamma c}(\chi_{u,i}-\chi_{d,i})\Gamma(1/3)\left(\frac{\omega}{4\omega_c}\right)^{-1/3}\left[\frac{\varphi_u^2-\varphi_d^2}{2}\right],
\label{eq:Ap1}
\end{aligned}
\ee
where $\Gamma(\nu)$ is the Gamma function.
In the scenario that $\varphi'\ll1/\gamma$ and $\chi'\ll1/\gamma$, only the imaginary component in Equation (\ref{eq:Ap1}) plays an important role on amplitude.
As shown in Figure \ref{fig:charges}, the limitations of $\varphi'$ read $\varphi_u=\varphi_t+\varphi$ and $\varphi_d=-\varphi_t+\varphi$.
Consequently, the total energy radiated per unit solid angle per unit frequency interval for charges in $i$th bunch is given by
\begin{equation}
\left.\frac{d^2W}{d \omega d \Omega}\right|_{i} \simeq\left(\frac{N_{\theta} N_{\phi}}{\Delta \theta_{s}}\right)^{2}\left(\chi_{u, i}-\chi_{d, i}\right)^{2} \frac{3 e^{2} \gamma^{2}}{4 \pi^{2} c}
\left\{
\begin{aligned}
&2^{2 / 3} \Gamma(2 / 3)^{2}\left(\frac{\omega}{\omega_{c}}\right)^{2 / 3}, ~& \omega \ll \omega_{c}, \\
&\pi\left(\frac{\omega}{\omega_{c}}\right) \exp \left(-\frac{\omega}{\omega_{c}}\right), ~& \omega \gg \omega_{c}.
\end{aligned}
\right.
\label{eq:spectruma}
\ee

For any $\chi'\gg1/\gamma\gg\varphi'$, the total amplitudes of the $i$th bunch projection on the plane perpendicular to a certain LOS can be written as
\be
\begin{aligned}
&A_{\|, i}\simeq\frac{2}{\sqrt{3}}\frac{\rho}{c}\frac{N_\theta}{\Delta\theta_s}\frac{N_\phi}{2\varphi_t}\int^{\chi_{u,i}}_{\chi_{d,i}}d\chi'\int^{\varphi_u}_{\varphi_d}\left[i\chi'^2K_{\frac{2}{3}}(\xi)+ \chi'|\chi'|K_{\frac{1}{3}}(\xi)\right]\cos\varphi'd\varphi',\\
&A_{\perp, i} \simeq \frac{2}{\sqrt{3}}\frac{\rho}{c}\frac{N_\theta}{\Delta\theta_s}\frac{N_\phi}{2\varphi_t}\int^{\chi_{u,i}}_{\chi_{d,i}}d\chi'\int^{\varphi_u}_{\varphi_d}|\chi'|K_{\frac{1}{3}}(\xi)\varphi'\cos\varphi'd\varphi'.
\end{aligned}
\label{eq:A2}
\ee
If $\omega\ll\omega_t\ll\omega_c$, one has $\xi\approx\omega\rho|\chi'|^3/3c\ll1$.
The amplitudes are same as Equation (\ref{eq:spectruma}).
If $\omega_t\ll\omega\ll\omega_c$, the parallel amplitude is given by
\be
A_{\|, i}\simeq(\varphi_u-\varphi_d)\frac{2}{\sqrt{3}}\frac{\rho}{c}\frac{N_\theta}{\Delta\theta_s}\frac{N_\phi}{2\varphi_t}\int^{\chi_{u,i}}_{\chi_{d,i}}\left(i\chi'^2+ \chi'|\chi'|\right)\sqrt{\frac{3c\pi }{\omega\rho|\chi'|^3}}\exp\left(-\frac{\omega\rho|\chi'|^3}{3c}\right)d\chi'.
\label{eq:Ap(omegallomega_t)}
\ee
Note that the error function is defined as
\begin{equation}
\mathrm {e r f}(x)=\frac{2}{\sqrt{\pi}} \int_{0}^{x} {\rm exp}(-\eta^{2}) d \eta.
\label{eq:error_function}
\end{equation}
According to Equation (\ref{eq:Ap(omegallomega_t)}) and the
properties of the error function, e.g., $\mathrm {e r f}(x)\rightarrow1$ for $x\rightarrow+\infty$, one has
\be
A_{\|, i}\simeq(\varphi_u-\varphi_d)\frac{i2\pi}{\sqrt{3}}\frac{1}{\omega}\frac{N_\theta}{\Delta\theta_s}\frac{N_\phi}{2\varphi_t}[\mathrm {e r f}(x_u)+\mathrm {e r f}(x_d)],
\ee
where $x_u=(\rho\omega|\chi_u|^3/3c)^{1/2}$ and $x_d=(\rho\omega|\chi_d|^3/3c)^{1/2}$.
$A_{\perp, i}\simeq0$ due to $\varphi\ll\chi'$.
The total energy radiated per unit solid angle per unit frequency interval for charges in $i$th bunch is given by
\begin{equation}
\left.\frac{d^2W}{d \omega d \Omega}\right|_{i} \simeq\left(\frac{N_{\theta} N_{\phi}}{\Delta \theta_{s}}\right)^{2}\left(\chi_{u, i}-\chi_{d, i}\right)^{2} \frac{3 e^{2} \gamma^{2}}{4 \pi^{2} c}
\left\{
\begin{aligned}
& \Gamma(2 / 3)^{2}\left(\frac{2\omega}{\omega_{c}}\right)^{2 / 3}+\frac{\gamma^{2}}{2^{2/3}}\left(\frac{\chi_{u, i}+\chi_{d, i}}{2}\right)^{2} \Gamma(1 / 3)^{2}\left(\frac{\omega}{\omega_{c}}\right)^{4 / 3}, ~& \omega \ll \omega_{t}, \\
&\left[\frac{4 \pi}{3 \gamma(\chi_{u, i}-\chi_{d, i})}\right]^{2}, ~& \omega_{t} \ll \omega .
\end{aligned}
\label{eq:spectrumb}
\right.
\ee

For $\varphi'\gg1/\gamma\gg\chi'$, we can just exchange $\chi_{u,i}$ and $\chi_{d,i}$ with $\varphi_u$ and $\varphi_d$ in Equation (\ref{eq:spectrumb}).
Therefore, the total energy radiated per unit solid angle per unit frequency interval for charges in the $i$th bunch is given by
\begin{equation}
\left.\frac{d^2W}{d \omega d \Omega}\right|_{i} \simeq\left(\frac{N_{\theta} N_{\phi}}{\Delta \theta_{s}}\right)^{2}\left(\chi_{u, i}-\chi_{d, i}\right)^{2} \frac{3 e^{2} \gamma^{2}}{4 \pi^{2} c}
\left\{
\begin{aligned}
&\Gamma(2 / 3)^{2}\left(\frac{2\omega}{\omega_{c}}\right)^{2 / 3}+\frac{\gamma^{2}}{2^{2/3}}\varphi^{2} \Gamma(1 / 3)^{2}\left(\frac{\omega}{\omega_{c}}\right)^{4 / 3}, ~& \omega \ll \omega_{t}, \\
&\left(\frac{2 \pi}{3 \gamma\varphi_t}\right)^{2} ~& \omega_{t} \ll \omega .
\label{eq:spectrumb2}
\end{aligned}
\right.
\end{equation}

Only if the size of the charge particle bunch is much smaller than the
half wavelength can the phase of radiation from the particles be approximately the same, which allows the emission to be highly coherent.
However, if $\omega\gg\omega_l$, the coherence of radiation would be reduced.
According to Equation (\ref{eq:B1}), the total energy radiated per unit solid angle per unit frequency interval for the moving bulk of bunches is given by
\be
\frac{d^2W}{d \omega d \Omega}=\sum_i^{N_s}\frac{e^{2} \omega^{2}}{4 \pi^{2} c}\left|-\boldsymbol{\epsilon}_{\|} A_{\|,i}+\boldsymbol{\epsilon}_{\perp} A_{\perp,i}\right|^2\simeq F_\omega(N_l, N_{lb})\frac{e^{2} \omega^{2}}{4 \pi^{2} c}\left|-\boldsymbol{\epsilon}_{\|} A_{\|,i}+\boldsymbol{\epsilon}_{\perp} A_{\perp,i}\right|^2,
\label{eq:sumi}
\ee
where $F_{\omega}(N_{l}, N_{lb})$ reads a dimensionless parameter denoting the enhancement factor due to coherence \citep{Yang18}:
\begin{equation}
F_{\omega}(N_l, N_{lb})=N_l^{2} N_{lb}^2\left[\frac{\sin \left(\pi l/\lambda \right)}{\left(\pi l/\lambda\right)}\right]^{2}\left[\frac{\sin \left(\pi l_{s}/\lambda\right)}{\left(\pi l_{s}/\lambda\right)}\right]^{2}.
\label{eq:Fomega}
\end{equation}
At higher frequencies, since oscillations become more rapid, the function could be approximated as a power law.
If $l_s\ll\lambda$, the total intensity is proportional to $N_l^2N_{lb}^2$.
If $l\ll\lambda\ll l_s$, the radiation from the bunches adds incoherently, i.e., $\mathcal{L}\propto N_{lb}N_{l}^2$.
Once $\lambda\ll l$ is satisfied, the radiation would become totally incoherent.

\section{Data analysis of FRB 190520}\label{app:190520}

In this section, we summarize the observation of an active newly discovered repeating FRB source, namely FRB 190520, by FAST \citep{Niu21}. The complex pulse morphology provides a way to reveal its intrinsic emission mechanism. 79 bursts have been analysed and 13 of them are found to have accompanied pulse(s) detected within 200 ms. Among the 13 bursts, 11 of them still emit at the ends of FAST frequency bandpass, so one is unable to measure their intrinsic frequency drift. We only keep the other 2 bursts for further calculations.

The dynamic spectral data has been de-dispersed first according to their best DM. To find the center frequency and the arrival time of each sub-burst, we made a Gaussian fit both on their spectrum and pulse profile. We take the peak value of Gaussian fit profile as the central frequency and arrival time. The pulse width and central frequency for the two sets of sub-pulses are [7.9, 11.4] ms, [1386, 1308] MHz and [9.8, 12.1] ms, [1376.8, 1332.7] MHz, respectfully. Then, we got the frequency drift rates  $-3.81\,\rm MHz\,ms^{-1}$ and $-2.79\,\rm MHz\,ms^{-1}$ for two sets of bursts from FRB 190520. Figure \ref{fig:dynamic190520} shows the frequency drift on the dynamic spectrum.

\begin{figure*}
\begin{center}
\includegraphics[width=0.45\textwidth]{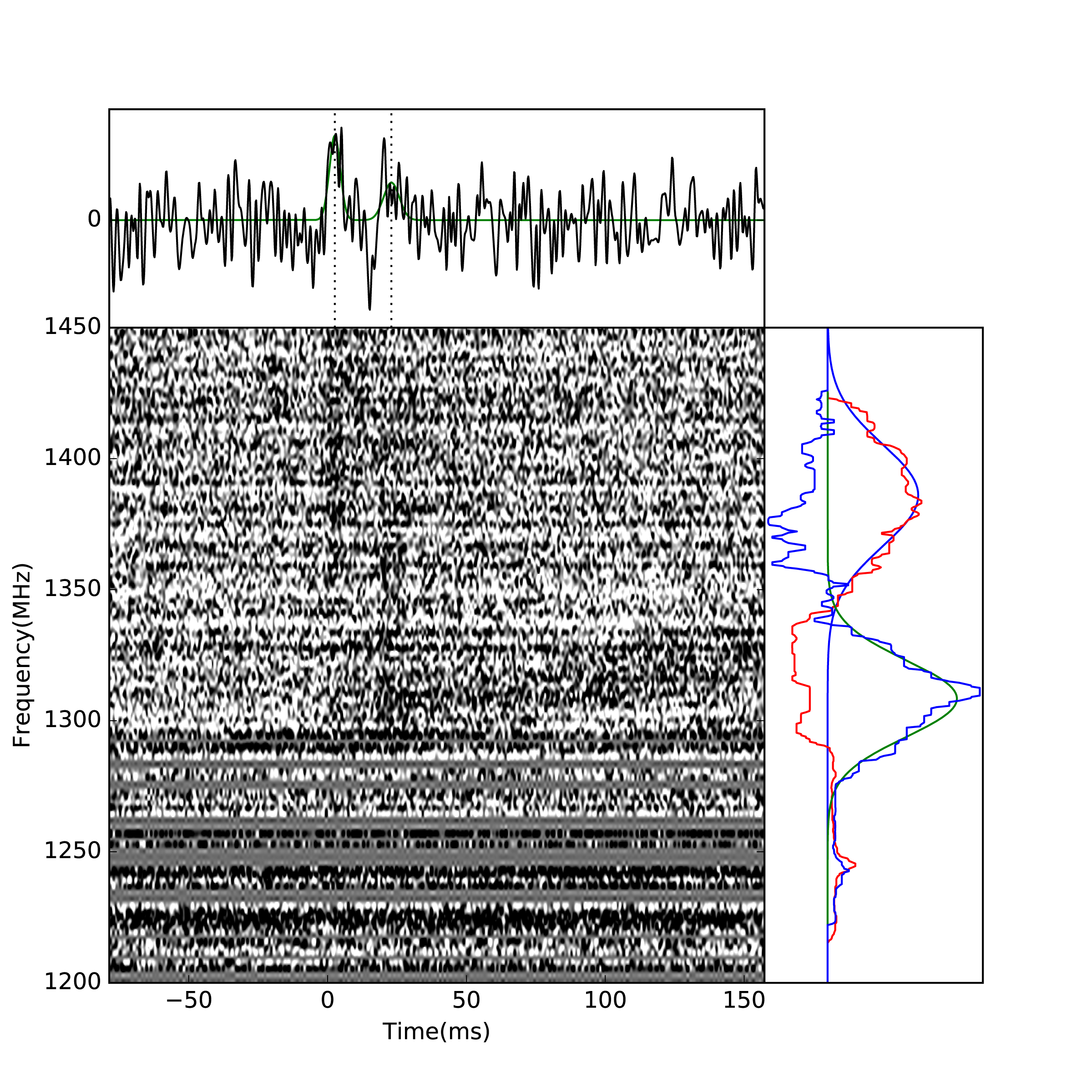}
\includegraphics[width=0.45\textwidth]{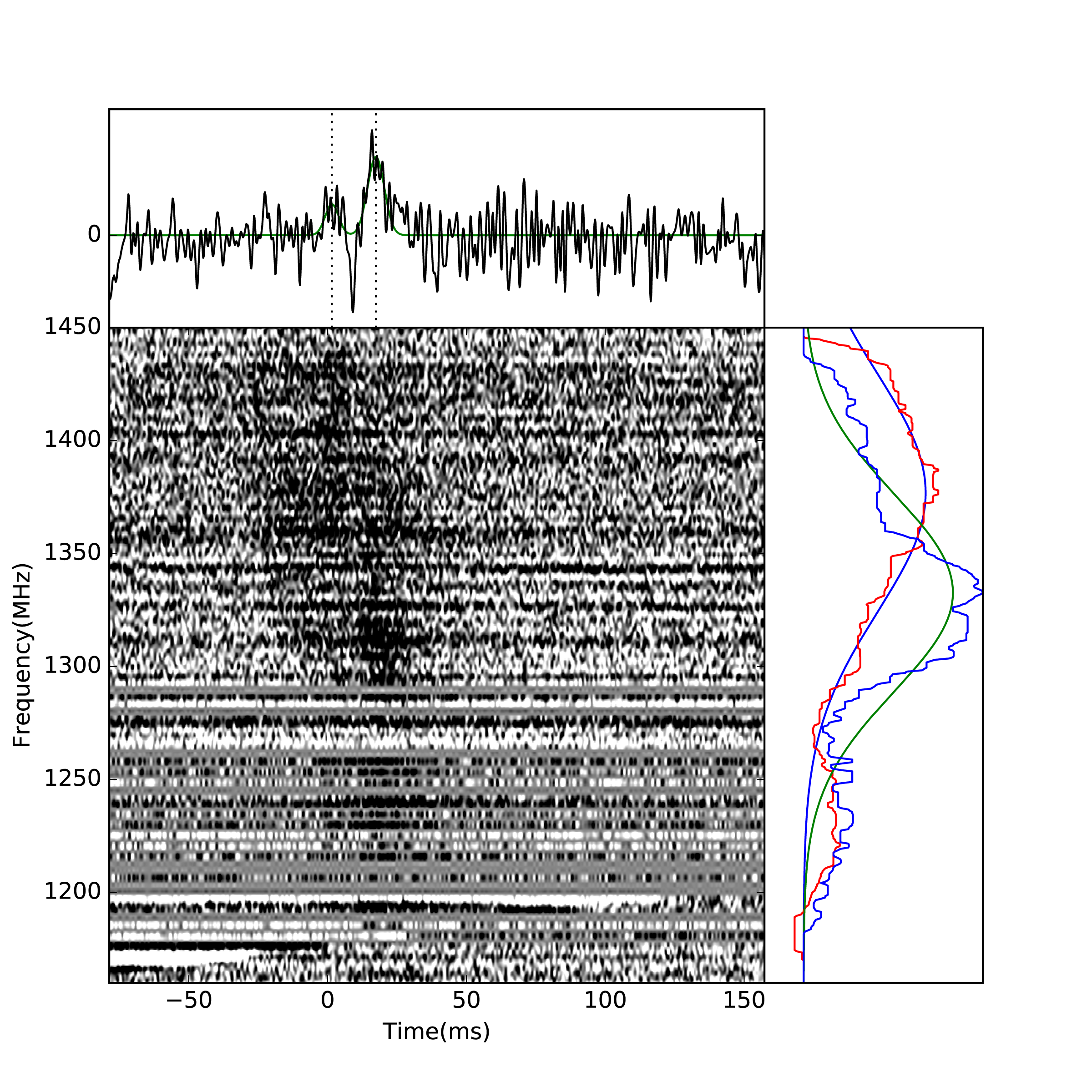}
\caption{\small{Dynamic spectra of two bursts from FRB 190520. The upper sub-plot shows the pulse intensity varying with time, including the two sub-pulses and its Gaussian-fit profile. The middle plot shows the dynamic spectrum after de-dispersion. The right plot shows the spectrum of each sub-pulse. }}
\label{fig:dynamic190520}
\end{center}
\end{figure*}

\clearpage

\section{Notation List}\label{app:notation}

\begin{tabular}{lll}
\hline
\hline
Symbol & Definition & First Appear\\
\hline
Subscript $i$, $j$, $k$ & The identifier of each charged particle & Section \ref{sec2.3}, Appendix \ref{app:coherent}\\
Subscript $x$ & $x$ is a number denoting order of magnitude & Section \ref{sec1}\\
$a$ & Strength parameter & Section \ref{sec2.2}\\
$c$ & Speed of light & Section \ref{sec2.1}, Appendix \ref{app:coherent}\\
$d^2W/d\omega d\Omega$ & Energy radiated per unit solid angle per unit frequency interval & Equation (\ref{eq:d2I/domegadOmega}), Appendix \ref{app:coherent}\\
$e$ & Elementary charge & Section \ref{sec2.1}, Appendix \ref{app:coherent}\\
$l$ & Length of a bunch & Section \ref{sec2.1}, Appendix \ref{app:coherent}\\
$l_s$ & Mean space between neighbourhood bunches & Section \ref{sec2.1}, Appendix \ref{app:coherent}\\
$m_e$ & Electron/positron mass & Section \ref{sec2.1}\\
$\boldsymbol{n}$ & Unit vector of the line of sight & Appendix \ref{app:coherent}\\
$n$ & A constant denoting the field configuration & Section \ref{sec2.1}, Appendix \ref{app:geometry}\\
$n_e$ & Number density of net charge & Equation (\ref{eq:ne})\\
$n_{\rm GJ}$ & Goldreich-Julian density & Equation (\ref{eq:nGJ})\\
$r$ & Distance to the neutron star center & Section \ref{sec2}\\
$s$ & Distance that charges traveled along the field lines & Section \ref{sec2.1}, Equation (\ref{eq:distance})\\
$t_{c}$ & Cooling timescale & Equation (\ref{eq:tcooling})\\
$t_{\rm int}$ & Burst intrinsic duration & Section \ref{sec2.1}\\
$v_e$ & Velocity of positron & Section \ref{sec2.1}, Appendix \ref{app:coherent}\\
$w$ & Sub-pulse width & Section \ref{sec2.1}\\
$z$ & Redshift & Section \ref{sec2.1}\\
$A$ & Flux function & Appendix \ref{app:geometry}\\
$A_{\parallel}$ & Parallel component of amplitude & Section \ref{sec2.3}, Appendix \ref{app:coherent}\\
$A_{\perp}$ & Perpendicular component of amplitude & Section \ref{sec2.3}, Appendix \ref{app:coherent}\\
$B$ & Magnetic field strength & Section \ref{sec2}\\
$B_s$ & Magnetic field strength at stellar surface & Section \ref{sec2.1}\\
$C_n(\theta)$ & A geometric factor & Equation (\ref{eq:curvatureradius0})\\
$I,\,Q,\,U,\,V$ & Stokes parameters & Section \ref{sec2.4}, Appendix \ref{app:coherent}\\
$\boldsymbol{E}$ & Wave electric field & Section \ref{sec2.1}\\
$E_{\parallel}$ & Electric field parallel to the B-field & Section \ref{sec2.1}\\
$F(A)$ & Free function & Section \ref{app:geometry}\\
$K_{\nu}$ & Modified Bessel function & Section \ref{sec2.3}, Appendix \ref{app:coherent}\\
$L$ Linear polarization & Section \ref{sec2.4}\\
$\mathcal{L}$ & Luminosity & Section \ref{sec2.1}\\
$\mathcal{L}_b$ & Luminosity of a bunch & Section \ref{sec2.1}\\
$\mathcal{L}_{\rm iso}$ & Isotropic equivalent luminosity in the observer frame & Equation (\ref{eq:Liso})\\
$p_e$ & Curvature radiation luminosity of a positron & Section \ref{sec2.1}\\
$\mathcal{M}$ & Multiplicity & Equation (\ref{eq:Multiplicity})\\
$N_b$ & Number of bunches contributing to instantaneous radiation & Section \ref{sec2.1}\\
$N_B$ & Total number of bunch & Section \ref{sec2.1}, Appendix \ref{app:coherent}\\
$N_e$ & Number of net charges in one bunch & Section \ref{sec2.1}, Appendix \ref{app:coherent}\\
$N_s$ & Maximum number of the subscript of $i$ & Section \ref{sec2.3}, Appendix \ref{app:coherent}\\
$N_\phi$ & Maximum number of the subscript of $k$ & Section \ref{sec2.3}, Appendix \ref{app:coherent}\\
$N_\theta$ & Maximum number of the subscript of $j$ & Section \ref{sec2.3}, Appendix \ref{app:coherent}\\
$P$ & Period of a neutron star & Section \ref{sec2.1}\\
\end{tabular}

\begin{tabular}{lll}
$P_{\rm gas}$ & Gas pressure & Equation (\ref{eq:Pgas})\\
$P_{\rm rad}$ & Radiation pressure & Equation (\ref{eq:Prad})\\
$R$ & Stellar radius & Section \ref{sec2.1}\\
$\mathcal{R}$ & Distance from the emitting source to the observer & Section \ref{sec2.3}\\
$T$ & Mean time interval between adjacent bunches & Section \ref{sec2.3}\\
$V_b$ & Volume of a bunch & Equation (\ref{eq:volume})\\
$\alpha$ & Angle between the magnetic axis and the rotational axis & Section \ref{sec2}\\
$\beta_e$ & Dimensionless velocity of a positron & Section \ref{sec2.3}, Appendix \ref{app:geometry}\\
$\gamma$ & Lorentz factor & Section \ref{sec2}, Appendix \ref{app:coherent}\\
$\lambda$ & Wave length & Section \ref{sec2.3}, Appendix \ref{app:coherent}\\
$\mu$ & A proportionality factor & Equation (\ref{eq:stokes})\\
$\nu$ & Frequency & Section \ref{sec2}\\
$\dot\nu$ & Drift rate & Section \ref{sec3.1}\\
$\chi$ & Angle between the electron velocity direction for different trajectories & Section \ref{sec2}, Appendix \ref{app:coherent}\\
$\chi_{d}$ & Lower boundary of $\chi$ & Section \ref{sec2.3}, Appendix \ref{app:coherent}\\
$\chi_{u}$ & Upper boundary of $\chi$ & Section \ref{sec2.3}, Appendix \ref{app:coherent}\\
$\psi$ & Polarization angle due to the rotation vectoer model  & Equation (\ref{eq:RVM})\\
$\rho$ & Curvature radius & Section \ref{sec2}, Appendix \ref{app:coherent}\\
$\sigma$ & Scattering cross section & Section \ref{sec2.2}\\
$\sigma_T$ & Thomson cross section & Section \ref{sec2.2}\\
$\sigma_w$ & Gaussian width & Equation (\ref{eq:gaussianprofile})\\
$\theta$ & Poloidal angle respect to the magnetic axis & Section \ref{sec2}, Appendix \ref{app:geometry}\\
$\theta_c$ & Spread angle of the curvature radiation & Equation (\ref{eq:thetac})\\
$\theta_{\rm jet}$ & Burst ``jet'' beaming angle & Section \ref{sec2.1}\\
$\theta_p$ & Angle between radial direction and tangent direction of magnetic field lines & Section \ref{sec3.1}, Equation (\ref{eq:thetap})\\
$\theta_s$ & Angle of the footpoint for field line at the stellar surface & Section \ref{sec2.1}\\
$\omega$ & Angular frequency & Section \ref{sec2.3}, Appendix \ref{app:coherent}\\
$\omega_B$ & Cyclotron frequency & Equation (\ref{eq:cyclotronfrequency})\\
$\omega_c$ & Critical angle frequency of curvature radiation & Section \ref{sec2.3}, Appendix \ref{app:coherent}\\
$\omega_l$ & Critical angle frequency for the bunch length & Equation (\ref{eq:omega_l}), Appendix \ref{app:coherent}\\
$\omega_t$ & Critical angle frequency for the trajectories & Section \ref{sec2.3}, Appendix \ref{app:coherent}\\
$\varphi$ & Azimuth angle respect to the magnetic
axis & Section \ref{sec2}\\
 & Angle between LOS and the trajectory plane & Appendix \ref{app:coherent}\\
$\varphi_t$ & Half open angle of the moving charges & Section \ref{sec2.1}, Appendix \ref{app:coherent}\\
$\zeta$ & Angle between the LOS and spin axis & Section \ref{sec2.4}\\
$\Gamma(\nu)$ & Gamma function & Appendix \ref{app:coherent}\\
$\Phi$ & Azimuth angle respect to the spin axis & Section \ref{sec2.4}\\
$\Phi_{\rm p}$ & Peak location of the Gaussian function & Equation (\ref{eq:gaussianprofile})\\
$\Psi$ & Polarization angle & Equation (\ref{eq:PA}),\\
$\Theta$ & The angle between magnetic axis and tangent direction of magnetic field lines & Equation (\ref{eq:Theta})\\
$\Omega$ & Angle frequency of a neutron star & Equation (\ref{eq:nGJ})\\
 & Solid angle of radiation & Section \ref{sec2}, Appendix \ref{app:coherent}\\
\hline \hline
\end{tabular}

\end{document}